%% file: paper_arxiv.tex
\documentclass[aps,pre,reprint,showpacs,notitlepage,onecolumn]{revtex4-1}
\usepackage{epsfig,graphics,amssymb,amsmath,subeqnarray,color,bm}
\usepackage[toc,page]{appendix}
\usepackage{mathtools}
\usepackage{xcolor}
\usepackage{float}

\numberwithin{equation}{section}
\usepackage{graphicx}
\usepackage[colorlinks=true,linkcolor=blue]{hyperref}

\input{definitions}

\begin{document}
\title{Maximizing propulsive thrust of a driven filament at low Reynolds number\\ via variable flexibility}
\author{Zhiwei Peng}
\affiliation{
Department of Mechanical Engineering,  
University of British Columbia,
Vancouver, B.C., V6T 1Z4, Canada}

\author{Gwynn J. Elfring}\email{Electronic mail: gelfring@mech.ubc.ca}
\affiliation{
Department of Mechanical Engineering,  
University of British Columbia,
Vancouver, B.C., V6T 1Z4, Canada}

\author{On Shun Pak}\email{Electronic mail: opak@scu.edu}
\affiliation{
Department of Mechanical Engineering,
Santa Clara University, Santa Clara, CA, 95053, USA }

\begin{abstract} 
At low Reynolds numbers the locomotive capability of a body can be dramatically hindered by the absence of inertia. In this work, we show how propulsive performance in this regime can be significantly enhanced by employing spatially varying flexibility. As a prototypical example, we consider the propulsive thrust generated by a filament periodically driven at one end. The rigid case leads to zero propulsion, as so constrained by Purcell's scallop theorem, while for uniform filaments there exists a bending stiffness maximizing the propulsive force at a given frequency; here we demonstrate explicitly how considerable further improvement (greater than 37\% enhancement compared with the optimal uniform stiffness) can be achieved by simply varying the stiffness along the filament. The optimal flexibility distribution is strongly configuration-dependent: while increasing the flexibility towards the tail-end enhances the propulsion of a clamped filament, for a hinged filament  decreasing the flexibility towards the tail-end is instead favorable. The results reveal new design principles for maximizing propulsion at low Reynolds numbers, potentially useful for developing synthetic micro-swimmers requiring large propulsive force for various biomedical applications.  
\end{abstract}
\maketitle

\section{Introduction}
\label{sec:intro}

For swimming microorganisms, such as spermatozoa or bacteria, the dominance of viscous forces over inertial effects leads to the time-reversible Stokes equations governing the fluid motion. In this low Reynolds number (Re) regime, Purcell's ``scallop theorem" states that a reciprocal motion (a deformation exhibiting time-reversal symmetry) cannot generate any net propulsive thrust \cite{purcell1977life}. In order to bypass the constraint of time-reversibility, some microorganisms in nature including flagellated bacteria and spermatozoa achieve self-propulsion by generating deformation waves along their flexible bodies \cite{brennen1977rev,Fauci06, berg2008coli,lauga2009hydrodynamics}. Meanwhile, advances in fabrication technologies at small scales have enabled the recent rapid development of synthetic micro-propellers capable of swimming at speeds comparable with these microorganisms \cite{yadav2015anatomy,Elgeti2015}. In particular, synthetic flexible filaments, such as nanowires and DNA linked with magnetic beads, have been employed to enable locomotion at small scales \cite{dreyfus2005microscopic,pak2011SoftMatter,williams2014Nature,Maier2016NANO}. While swimming microscale robots have the potential to carry out minimally invasive medical operations \cite{Ebbens2010,nelson2010microrobots,Nelson2014}, improving the swimming efficiency and propulsive performance will be important aspects of future designs.

Different physical mechanisms that can enable or enhance propulsion at low Re have been considered, including the presence of boundaries \cite{Tierno2008, Sing2010, Zhang2010, zhu2013, Aguilar2013}, heterogeneous environments \cite{Leshansky2009, Fu2010}, viscoelasticity \cite{Teran2010, Liu2011, Nathan2012, Garcia2013, Spagnolie2013, Riley2014, Thomases2014, wrbel2016}, and shear-dependent viscosity \cite{Johnson2013, qiu2014swimming, Li2015, Datt2015, Park2016}. In particular, propulsion as a result of the interplay between hydrodynamics and elasticity has been extensively studied \cite{Machin1958, Wiggins1998, WigginsPRL1998, Lowe2003, Manghi2006, Gauger2006, Lauga2007PRE, Coq2008, Keaveny08, Fu2008, Qian2008}. For instance, a rigid filament driven at one end cannot propel itself because the motion is reciprocal, but by introducing flexibility in the filament, the coupling between viscous and elastic forces produces deformation along the filament that can lead to propulsion \cite{Wiggins1998,WigginsPRL1998}.  For a given actuation frequency and filament length, an optimal bending stiffness of the filament can be determined to produce the largest propulsive force \cite{Machin1958, Wiggins1998, WigginsPRL1998, lauga2009hydrodynamics}; however, the possibility of further improving the propulsion by allowing variable flexibility along the filament remains largely unexplored.  

The interaction of flexible bodies with fluid flows at high Reynolds number has long been studied \cite{shelley11, dewey2013} and considerable attention has been on the role of flexibility in animal locomotion \cite{wu11}. Interestingly, some flying animals, such as hoverflies and hummingbirds, exhibit a nonuniform distribution of flexibility along their wings, which can potentially enhance the propulsive performance \cite{tanaka2011effect,song2015}. For a flapping wing in this high Reynolds number regime, Shoele and Zhu \cite{shoele2013pof} have compared the performance of several cases of nonuniform flexibility distributions, while a recent study by Moore \cite{moore2015torsional} showed that optimal propulsion can be achieved by having highly localized flexibility at the front of a wing using a torsional spring arrangement. To investigate the effect of non-uniform bending stiffness on fish swimming, Lucas \textit{et. al.} \cite{lucas2015effects} constructed foils of segmented stiffness profiles which are experimentally shown to enhance the swimming performance. At low Reynolds number, the exploration of enhanced propulsion via variable flexibility is only yet nascent. Maier \textit{et.~al.~}\cite{Maier2016NANO} constructed an artificial micro-swimmer with a flagellar bundle possessing an exponentially decreasing stiffness profile by using a DNA tile-tube assembly attached to a magnetic head. It was found that the micro-swimmers with exponentially decreasing stiffness could outperform swimmers with uniform stiffness along the flagellum under a rotational actuation. 

The above pioneering studies prompt an array of fundamental questions: What stiffness profiles aside from exponential distributions can further enhance propulsion? Does the optimality of the torsional spring arrangement at high Reynolds numbers still hold in the low-Reynolds-number world? Are there any general design principles for maximizing the propulsive thrust with non-uniform flexibility? In this work, we address these questions by considering the propulsive thrust generated by a slender filament driven at one end at low Re. We demonstrate explicitly how spatially varying flexibility can significantly enhance the propulsive thrust by exploring a reduced-space of basis functions for the flexibility distributions. 

The paper is organized as follows. In Sec.~\ref{sec:formulation} a general formulation for the elastohydrodynamics of a boundary-actuated flexible filament immersed in a viscous fluid is presented, before considering asymptotic expansions for the solution of the nonlinear problem in the small-amplitude oscillation limit.  
In Sec.~\ref{sec:resDiscus} we first show that though optimal at high Reynolds number, a torsional spring arrangement cannot outperform the optimal uniform stiffness at low Reynolds number. We then demonstrate that, for a cantilevered filament, a higher propulsive thrust can be achieved by having a decreasing stiffness from the driven end. However, we emphasize in Sec.~\ref{sec:torque_free} that the conclusion cannot be simply generalized to other mechanical tethering conditions. As an example, we show that a hinged filament instead favors an increasing stiffness from the driven end. Finally we conclude the work with remarks in Sec.~\ref{sec:conclude}.

\section{Problem Formulation}
\label{sec:formulation}

\subsection{Elastohydrodynamics}
\begin{figure}[!tb]
\centering
\includegraphics[width=0.45\textwidth]{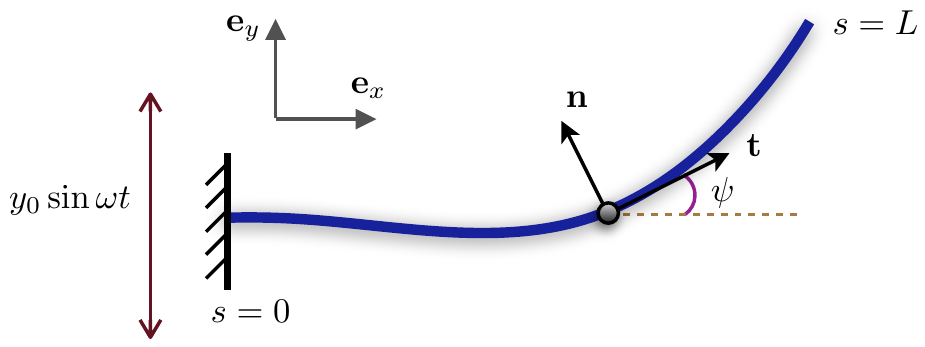}
\caption{\label{fig:schematic}Schematic diagram of a displacement-driven cantilevered filament and notation.}
\end{figure}

We consider an elastic and inextensible filament of length $L$ and radius $r$. The filament is slender, $r/L \ll 1$, and is immersed in a viscous fluid such that the Stokes equations apply. The position vector of a material point on the filament neutral line relative to the laboratory frame at time $t$ is defined as $\bx(s,t)$, where $s \in [0, L]$ is the arclength along the filament. We define the local unit tangent and normal vectors as $\bt = \bx_s = \cos\psi\be_x+\sin\psi\be_y$ and $\bn = \be_z\times \bt$, where $\psi(s,t)$ is the angle between the tangent vector $\bt$ and the positive $x$-direction ($\be_x$) and the subscript $s$ denotes differentiation with respect to the arclength (see Fig.~\ref{fig:schematic} for a schematic). 

The elastic filament is modeled as an Euler-Bernoulli beam with an energy functional \cite{Landau,Goldstein1995PRL}
\begin{align}
\mathcal{E} = \frac{1}{2}\int_0^L A\bx_{ss}^2\diff s+\frac{1}{2}\int_0^L\sigma\left(  \bx_s\cdot\bx_s-1\right)\diff s,
\end{align}
where the bending stiffness (or rigidity), $A(s)$, is allowed to vary along the filament.
The local inextensibility condition, $ \bx_s\cdot\bx_s=1$, is enforced by the Lagrange multiplier $\sigma(s,t)$. The elastic force density along the filament can then be obtained by a variational derivative, 
\begin{align}
\bf_\text{elastic} = - \delta\mathcal{E}/\delta \bx =- \partial_s\left[ \partial_s(A\kappa)\bn-\tau\bt \right],
\end{align}
where the tensile force along the filament $\tau = \sigma+A\kappa^2$ and the local curvature $\kappa$ can be calculated as $\kappa = \lVert\bx_{ss}\rVert =\psi_s$. In contrast to previous studies that assume a uniform bending rigidity \cite{Goldstein1995PRL,camalet1999PRL,camalet2000,Yu2006, Lauga2007PRE,Evans2010PRE,spagnolie2010optimal,pak2011SoftMatter,rk_HFSP_J}, we emphasize that here the bending rigidity is permitted to vary spatially. 

The hydrodynamics of slender bodies at low Reynolds number is described by the resistive force theory (RFT) to leading order \cite{Gray1955, lighthill1975mathematica,childress1981mechanics}, where the viscous force per unit length on the body is linearly related to its local velocity: 
\begin{align}
\bf_\text{vis} = -\left( \xi_\perp\bn\bn+\xi_\parallel\bt\bt \right)\cdot\bx_t,\label{eq:fvisc}
\end{align}
where  $\xi_\parallel$ and $\xi_\perp$ are the resistive coefficients in the tangential and normal directions respectively. Although non-local hydrodynamic interactions are ignored in RFT, this local theory has often proved effective in obtaining predictions in quantitative agreement with experiments \cite{Gray1955, Wiggins1998,Yu2006, dreyfus2005microscopic, pak2011SoftMatter}.

Neglecting the inertia of the filament there is a local balance between the viscous and elastic forces, 
\begin{align}\label{eq:forceBalance}
\bf_\text{vis}+\bf_\text{elastic} = \mathbf{0},
\end{align}
which together with the local inextensibility condition result in a set of coupled nonlinear partial differential equations governing the evolution of the filament shape, $\bx(s,t)$, in terms of tangent angle $\psi(s,t)$ and tensile force $\tau(s,t)$:
\begin{align}
&\psi_t =\frac{1}{\xi_\perp}\left[ -\partial_s^3(A\psi_s)+\partial_s(\psi_s\tau) \right]+\frac{1}{\xi_\parallel}\psi_s\left[ \psi_s\partial_s(A\psi_s)+\tau_s \right],\label{eq:govern1}\\
&\tau_{ss}-\frac{\xi_\parallel}{\xi_\perp}\psi_s^2\tau=-\partial_s(\psi_s\partial_s(A\psi_s))-\frac{\xi_\parallel}{\xi_\perp}\psi_s\partial_{ss}(A\psi_s)\label{eq:govern2}.
\end{align}

\subsection{Boundary actuation}
We consider here a filament where one end ($s=0$) is oscillated harmonically 
\begin{align}
y(0,t)=y_0\sin\omega t,
\end{align}
with $\omega$ denoting the oscillation frequency \cite{Wiggins1998, WigginsPRL1998, Yu2006,Evans2010PRE, moore2015torsional}. The elastohydrodynamic response of the filament varies depending on the tethering mechanism of the filament, whether it is cantilevered: $\psi(0,t)=0$ (Secs.~\ref{subsec:conti}-\ref{subsec:two_seg}), hinged: $\psi_s(0,t)=0$ (Sec.~\ref{sec:torque_free}), or connected to other devices  such as a torsional spring (Sec.~\ref{subsec:torsional}). These different scenarios at the actuation end will be analyzed in subsequent sections. 

At the free end ($s=L$), the filament is force-free and torque-free:
\begin{align}\label{eq:ForceTorqueBC}
\bF_\text{ext}(L) &= [\tau\bt-\partial_s(A\kappa)\bn]_{s=L}=\mathbf{0}, \nonumber\\
\quad T_\text{ext}(L) &= [A\kappa]_{s=L}= 0,
\end{align}
where the expressions for external forces and torques arise as boundary terms in the variational calculation \cite{audoly2010elasticity}. The propulsive force generated by the boundary actuation can be obtained by integrating the viscous force along the filament averaged over a period of actuation,
\begin{align}
F_\text{p} = - \left< \be_x\cdot\int_0^L \bf_\text{vis}\diff s \right>,
\end{align}
where the angle brackets represent time averaging over an oscillation period.

We consider here small amplitude oscillations, $\epsilon = y_0/L \ll 1$, where deformations may be assumed to be periodic and we determine the filament shape asymptotically, order by order.  For small amplitude oscillations, we write the net propulsive force as a regular series expansion in $\epsilon$, $F_\text{p} = \epsilon^2F_\text{p}^{(2)}  + \mathcal{O}(\epsilon^4)$ (the odd powers vanish due to the $\epsilon \to -\epsilon$ symmetry) and upon substitution of Eq.~\eqref{eq:fvisc}, assuming a continuous stiffness $A(s)$, we obtain
\begin{align}\label{eq:prop_general}
F_\text{p}^{(2)} = \frac{\xi_\perp-\xi_\parallel}{\xi_\perp}\left<\int_0^L \psi^{(1)}\partial_s^2(A\partial_s\psi^{(1)}) \diff s \right>,
\end{align}
where $\psi(s,t) = \epsilon\psi^{(1)}+\mathcal{O}(\epsilon^2)$.

We now non-dimensionalize the governing equations with respect to a length scale $L$, time scale $\omega^{-1}$ and a force scale $L^2\xi_\perp \omega$. Dimensionless groups obtained include the dimensionless oscillation amplitude $\epsilon = y_0/L$, the drag anisotropy ratio $\gamma = \xi_\perp/\xi_\parallel$, and the sperm number $\text{Sp}=L (\xi_\perp \omega/A_0)^{1/4}$ where $A_0$ is a characteristic magnitude of the bending stiffness. Unless otherwise noted we scale stiffness by the value at the driven end, $A_0=A(0)$, and introduce a dimensionless bending stiffness $\Af(s) = A(s)/A(0)$.  In the limit of an infinitely slender filament, $L/r\to \infty$, the drag anisotropy ratio $\gamma = 2$, which is the value adopted in this work. The sperm number Sp compares the magnitude of characteristic viscous and elastic forces. For a given oscillation frequency $\omega$ and length of the filament $L$, a larger $\text{Sp}$ indicates a more flexible material and $\text{Sp}\to 0$ tends to a rigid filament.
Hereafter, we shall work with dimensionless variables while adopting the same notation introduced above.

\section{Results and Discussion}
\label{sec:resDiscus}
For the classical case of a uniform bending stiffness along the filament \cite{Machin1958, Wiggins1998, WigginsPRL1998, lauga2009hydrodynamics}, $\Af=1$, the propulsive force varies non-monotonically as a function of the sperm number as shown in Fig.~\ref{fig:torsional} (red dashed line). At low Sp, the filament tends to a rigid rod undergoing a reciprocal motion that produces zero net propulsive thrust. For finite values of Sp, flexibility enables the propagation of deformation waves along the body and the generation of a net propulsive force. At large Sp, the dominance of viscous forces over elastic forces suboptimally localizes filament deformation around the actuation end, leaving a considerable portion of the filament undeformed and not contributing to propulsion. The propulsive force hence exhibits a maximum around an intermediate sperm number $\text{Sp}\approx 1.89$, leading to a maximum propulsive thrust of $F_\text{p}^{(2)} = 0.19625$.

\subsection{Localized flexibility: via a torsional spring}
\label{subsec:torsional}
Before studying various spatially varying flexibility distributions in later sections, we first investigate a localized flexibility arrangement via a torsional spring, which was reported to optimize propulsive thrust at high Reynolds numbers \cite{moore2015torsional}. We ask the question: is torsional spring also the optimal flexibility arrangement at low Reynolds number? 

We consider a rigid filament connected by a torsional spring with a spring constant $C$ at the actuation end. In this case the local tangent remains the same along the rigid filament, $\psi(t)= \epsilon \psi^{(1)}+ \mathcal{O}(\epsilon^2)$, and the torque balance of the entire filament reads
\begin{align}
\be_z\cdot\int_0^1[\bx(s,t)-\bx(0,t)]\times \bf_\text{vis}\diff s - K \psi=0, 
\end{align}
where $K =C/{L^3\xi_\perp\omega}$ is the dimensionless spring constant. Similar to the sperm number Sp, the dimensionless spring constant $K$ measures the magnitude of the elastic force compared with the viscous force. The above torque balance exactly simplifies to $K\psi + \psi_t/3+ \epsilon \cos\psi\cos t/2=0$. At leading order in $\epsilon$, we have
\begin{align}
\frac{1}{3}\psi_t^{(1)}+\frac{1}{2}\cos t +K\psi^{(1)}=0,
\end{align}
with $\psi^{(1)}(0)=0$. The periodic (or long time) solution is given by
\begin{align}
\psi^{(1)}(t) = -\frac{3}{2+ 18 K^2}\left( 3 K\cos t+\sin t\right).
\end{align}

\begin{figure}[!tb]
\centering
\includegraphics[width = 0.45\textwidth]{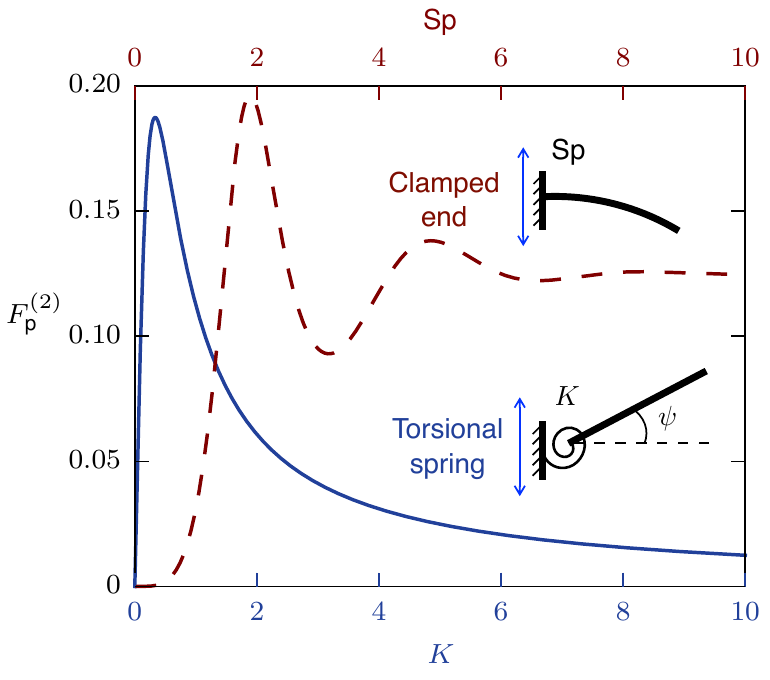}
\caption{\label{fig:torsional}Red: propulsive force generated by a displacement-driven cantilevered filament of uniform bending stiffness. The maximum propulsive force here is $F_\text{p}^{(2)} = 0.19625$. Blue: propulsive force generated by a torsional spring arrangement at the actuation end connected to a rigid filament as a function of the spring constant. The maximum propulsive force here is $F_\text{p}^{(2)} = 0.1875$.}
\end{figure}
Using this result we obtain a simple analytical expression for the dimensionless propulsive force, 
\begin{align}
F_\text{p}^{(2)} = \frac{\gamma-1}{\gamma} \frac{9K}{4(1+9K^2)}.
\end{align}
The non-monotonic variation of the propulsive force as a function of the dimensionless spring constant is shown in Fig.~\ref{fig:torsional} (blue solid line) and the optimal propulsive force $F_\text{p}^{(2)} = 0.1875$ occurs at $K=1/3$. These results can be understood by considering the horizontal force (propulsive thrust) that results from vertically moving a rigid filament inclined at an angle $\psi$, which scales as $f_\text{prop} \sim \cos\psi\sin\psi$ to leading order \cite{lauga2009hydrodynamics}, while the extrema of the thrust occur at $\psi= \pm45^\circ$, which implies that the more time a rigid filament is inclined around these optimal angles during an oscillation the larger the resulting net propulsive force. For the torsional spring arrangement, the average tangent angle over a peak-to-peak stroke is given by $9 K/(\pi (1 + 9K^2))$, which has a maximum of $3/(2\pi)$ ($\approx27^\circ$) occurring at $K=1/3$. This explains the occurrence of the optimal propulsive force at $K=1/3$ and as $K$ deviates from this optimal value, the average tangent angle decreases leading to suboptimal propulsion. 

However, we emphasize that the maximum possible propulsive thrust obtained via this torsional spring arrangement ($F_\text{p}^{(2)} = 0.1875$) is indeed smaller than that given by the optimal uniform stiffness configuration ($F_\text{p}^{(2)} = 0.19625$, see Fig.~\ref{fig:torsional} for comparison). Therefore we obtain the conclusion: despite the superiority of the torsional spring arrangement in high Reynolds number propulsion \cite{moore2015torsional}, such a localized flexibility arrangement is suboptimal at low Reynolds number compared with the optimal uniform stiffness.

\subsection{Continuous stiffness distributions}
\label{subsec:conti}

In this section we explore the possibility of maximizing the propulsive performance with a continuously variable stiffness. Using Eqs.~(\ref{eq:govern1}) and (\ref{eq:govern2}), at order $\epsilon$ we have
\begin{align}
\text{Sp}^4 \partial_t\psi^{(1)}+\partial_s^3\left( \Af \partial_s\psi^{(1)} \right)=0. \label{eqn:firstorderdimensionless}
\end{align}
Assuming periodic solutions,
\begin{align}
\psi^{(1)} = \mathfrak{R}\{ e^{it} h(s)\},
\end{align}
where $\mathfrak{R}$ indicates the real part of a complex quantity, we numerically solve the resulting ODE for $h(s)$, from which the propulsive force can be evaluated using Eq.~(\ref{eq:prop_general}) with numerical integration for a given stiffness profile $\Af(s)$ (see Sec.~\ref{subsec:continuous_num} for details). 

Following the studies by Moore \cite{moore2015torsional} in the high Reynolds number regime, we optimize the propulsive force with a linear and quadratic distribution of bending stiffness. We also explore an exponentially decreasing stiffness distribution as considered by Maier \textit{et al.} \cite{Maier2016NANO} at low Reynolds number.

\begin{figure}[!tb]
\centering
\includegraphics[width = 0.45\textwidth]{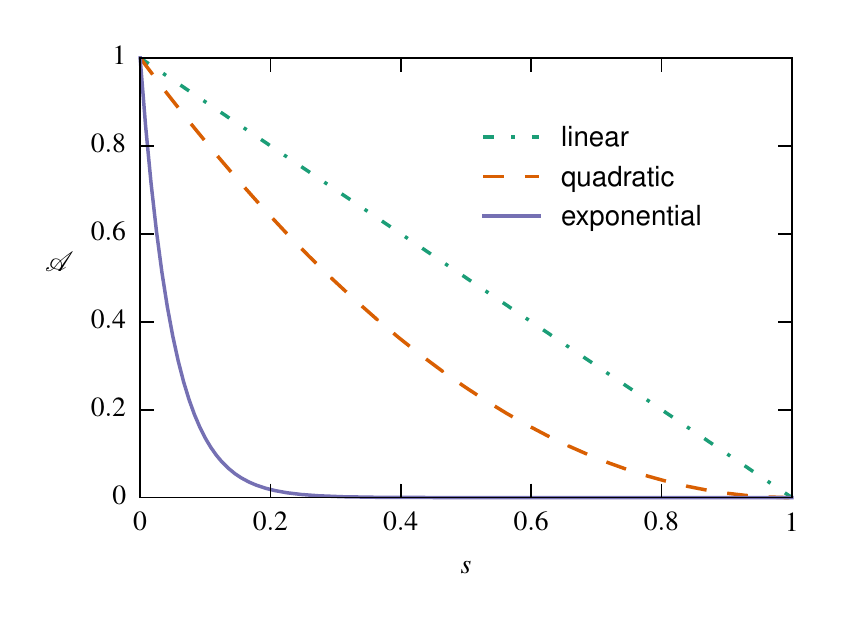}
\caption{\label{fig:continuous} Optimal linear ($\text{Sp}\approx 1.77, a =0, b= -1, F_\text{p}^{(2)}\approx 0.2012$), quadratic ($\text{Sp} \approx 1.64, a = 1, b=-2, F_\text{p}^{(2)}\approx 0.2124$)  and exponential stiffness distributions.}
\end{figure}

For both linear and quadratic distributions, given by $\Af(s) = as^2+bs+1$, we optimize numerically over the parameter space $(\text{Sp}, a, b)$ with the constraint of positivity: $\text{Sp}>0$, $\Af(s)>0$ for any $s\in[0,1)$. In both cases the optimization routine points to a limiting scenario of a vanishing bending stiffness at the free end (see Sec.~\ref{subsec:opt} for details). The optimal linear and quadratic distributions are shown in Fig.~\ref{fig:continuous}, which improve the propulsive performance respectively by 2.5\% and 8.2\% compared with the optimal uniform stiffness. We also find that an optimal exponentially decaying stiffness,  $\Af(s) = \text{exp}(a s)$, can further increase the propulsive thrust by 13.5\% compared with the optimal uniform stiffness case in our displacement-driven boundary setup. The optimal exponential profile ($\text{Sp}\approx 0.1, a \approx-20$) is shown in Fig.~\ref{fig:continuous}. All these optimal distributions display the advantage of decreasing stiffness from the driven end to enhance propulsion under the current boundary actuation.

\subsection{Segmented stiffness distributions}
\label{subsec:two_seg}

\begin{figure*}[!tb]
\centering
\includegraphics[width =0.9\textwidth]{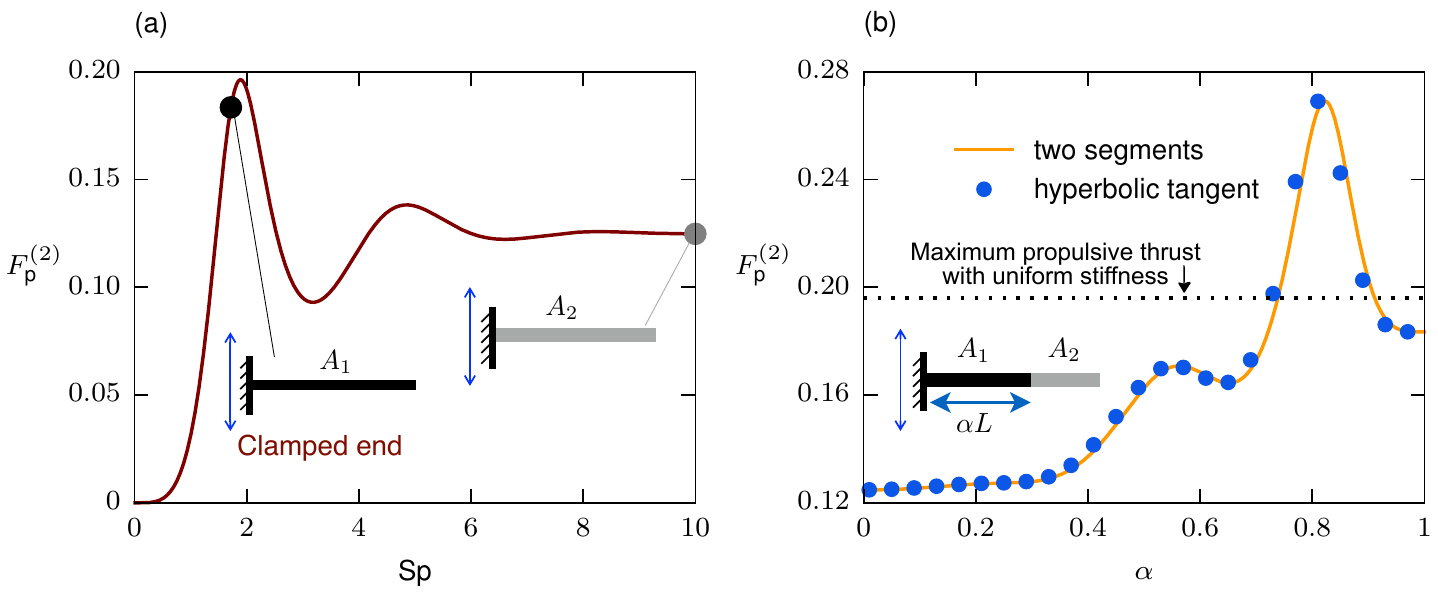}
\caption{\label{fig:uni_2seg} (a) Propulsive thrust generated by a cantilevered filament of uniform bending stiffness under a  displacement actuation. The two dots (black and gray) indicate two filaments with different bending stiffnesses $A_1$ and $A_2$, or $\text{Sp}= 1.7189$ and $\text{Sp}=10$.  (b) Propulsive thrust generated by a two-segment filament with $\text{Sp}_1= 1.7189$ and $\text{Sp}_2=10$ as a function of $\alpha$. The maximum propulsive force generated by this segmented stiffness profile is 37\% greater than the maximum achievable thrust of a filament with any uniform stiffness. The blue dots indicate numerical results using a regularized hyperbolic tangent function that approximates a step-function stiffness distribution.}
\end{figure*}

We have demonstrated that for a displacement-driven actuation at the cantilevered end of the filament, allowing the bending stiffness to vary continuously along the entire filament leads to improved propulsive performance compared with any uniform stiffness. Despite the observed enhancement, the polynomial and exponential distributions studied may not necessarily be ideal candidates to begin with. We explore in this section stiffness distributions with jump discontinuities,  \textit{i.e.}, a step-function distribution, for enhancing the propulsive thrust because such segmented flexibility arrangements may be easily implemented experimentally by simply serially connecting filaments with different stiffnesses as diagrammed in Fig.~\ref{fig:uni_2seg}(b). Recently, fish-like foil models with segmented flexibility distributions have also been constructed and experimentally shown to enhance swimming at high Reynolds number \cite{lucas2015effects}.

We assume different flexibilities for the segments at the actuation end, $A_1$, and the distal end, $A_2$, with the ratio of stiffness denoted as $\beta = A_2/A_1$. The length of the segment with stiffness $A_1$ relative to that of segment with stiffness $A_2$ is denoted as  $\alpha \in [0,1]$ (Fig.~\ref{fig:uni_2seg}(b)).

In this case, the local tangent $\psi(s,t)$ and tensile force $\tau(s,t)$ are split into two separate functions for the two segments, with $\psi_k^{(n)}$ denoting the tangent angle of the $k$-th segment ($k=1,2$) at $\mathcal{O}(\epsilon^n)$. Noting that $\psi \sim \epsilon$, from Eqs.~(\ref{eq:govern1}) and (\ref{eq:govern2}) one can show that $\tau\sim \epsilon^2$ (see Ref.~\cite{pak2014theoretical}). If both $A_1$ and $A_2$ are finite, the leading order equations of motion are written as
\begin{align}
\text{Sp}_1^4\partial_t\psi_1^{(1)} +\partial_s^4\psi_1^{(1)}&=0,\label{eq:psi1}\\
\text{Sp}_2^4\partial_t\psi_2^{(1)} +\partial_s^4\psi_2^{(1)}&=0,\label{eq:psi2}
\end{align}
where Sp$_1=L\left( \xi_\perp\omega/A_1 \right)^{1/4}$ is the sperm number of the material at the driven end,  Sp$_2=L\left( \xi_\perp\omega/A_2 \right)^{1/4}$ is the sperm number of the material at the free end, and $\beta = (\text{Sp}_1/\text{Sp}_2)^4$. Again we take $\psi_k^{(1)} = \mathfrak{R}\{h_k(s)e^{it}\}$,  and solve the corresponding ordinary differential equations for $h_k(s)$. 

The boundary conditions at the two ends are given by
\begin{align}
&\psi_1^{(1)}(0,t)=0, \quad \partial_s^3\psi_1^{(1)}(0,t) = -\text{Sp}_1^4\cos t,\nonumber\\
&\partial_s\psi_2^{(1)}(1,t)=0, \quad \partial_s^2\psi_2^{(1)}(1,t)=0.
\end{align}
At the connecting point, the continuity of tangent angle, internal force and torque result in the conditions
\begin{align}
&\psi_1^{(1)}(\alpha,t)=\psi_2^{(1)}(\alpha,t), \quad \partial_s\psi_1^{(1)}(\alpha,t) = \beta\partial_s\psi_2^{(1)}(\alpha,t)\nonumber\\
&\partial_s^2\psi_1^{(1)}(\alpha,t) = \beta\partial_s^2\psi_2^{(1)}(\alpha,t), \quad \partial_s^3\psi_1^{(1)}(\alpha,t) = \beta\partial_s^3\psi_2^{(1)}(\alpha,t).
\end{align}
After separation of variables the resulting ordinary differential equations for $h_k$ can be solved by 
\begin{align}
h_k(s) = \sum\limits_{n=1}^{4}c_n \text{exp}(\lambda_n s),
\end{align}
where $k=1, 2$ and the constants $c_n$ can be solved by imposing the boundary conditions above.

Once the deformation along the segments is obtained, the leading order propulsive force is then given by
\begin{align}\label{eq:prop_2segCant}
F_\text{p}^{(2)} = \frac{\gamma-1}{2\gamma\text{Sp}_1^4}\biggl< (\partial_s\psi_1^{(1)}(0,t))^2  - (\partial_s\psi_1^{(1)}(\alpha,t))^2+\beta (\partial_s\psi_2^{(1)}(\alpha,t))^2 \biggr>.
\end{align}
Now, inspired by the findings in previous sections, we probe the possibility of enhancing propulsion by connecting a more flexible segment at the free end to the segment at the actuation end ($\beta<1$).  As shown in Fig.~\ref{fig:uni_2seg}, varying the relative proportion of two segments with distinct flexibilities leads to a non-monotonic variation of the propulsive force. The parameters presented in Fig.~\ref{fig:uni_2seg} are the optimal values obtained by searching over the parameter space  $\text{Sp}_1,\ \text{Sp}_2 \in (0,10]$ and $\alpha \in [0, 1]$. In this case the limits $\alpha=0$ and $\alpha=1$ reduce to the cases of uniform stiffness with $\text{Sp} = 10$ and $\text{Sp} = 1.7189$ respectively. By tuning the relative portion of the two segment $\alpha$, one can obtain an optimal ratio $\alpha = 0.8224$ generating a propulsive thrust that is 37\% better than the optimal uniform case, an improvement not achievable by any continuous stiffness distribution considered previously. We emphasize here that segmented flexibility distributions maximizing propulsion also have decreasing stiffness distributions, consistent with the previously observed trend.

As a remark, the segmented distribution (or step-function distribution) may be reconstructed by a continuous hyperbolic tangent distribution: $\Af(s)=a+b \tanh \left[ (s-\alpha)/\delta \right]$, where $\delta \ll 1$ is a parameter regularizing the sharp transition between the two distinct bending stiffnesses. The parameters $a, b$ are determined by satisfying $\Af(0)=1$, $\Af(1) = \beta$. Using the formulation in Sec.~\ref{subsec:conti} and the regularization parameter $\delta = 0.003$, one obtains effectively the same results (represented by blue dots in Fig.~\ref{fig:uni_2seg}b) as the segmented distribution. 

One may improve the propulsion of a two-segment elastic filament by searching over a wider parameter regime or even by looking at optimal three-segment and four-segment elastic filaments but we note that this results in exceedingly small and flexible pieces at the end, unrealistic for experimental fabrication and severely constraining our assumption of small amplitude deformations.

\section{Effects of boundary actuation mechanisms}
\label{sec:torque_free}

\begin{figure*}[!tb]
\centering
\includegraphics[width =0.9\textwidth]{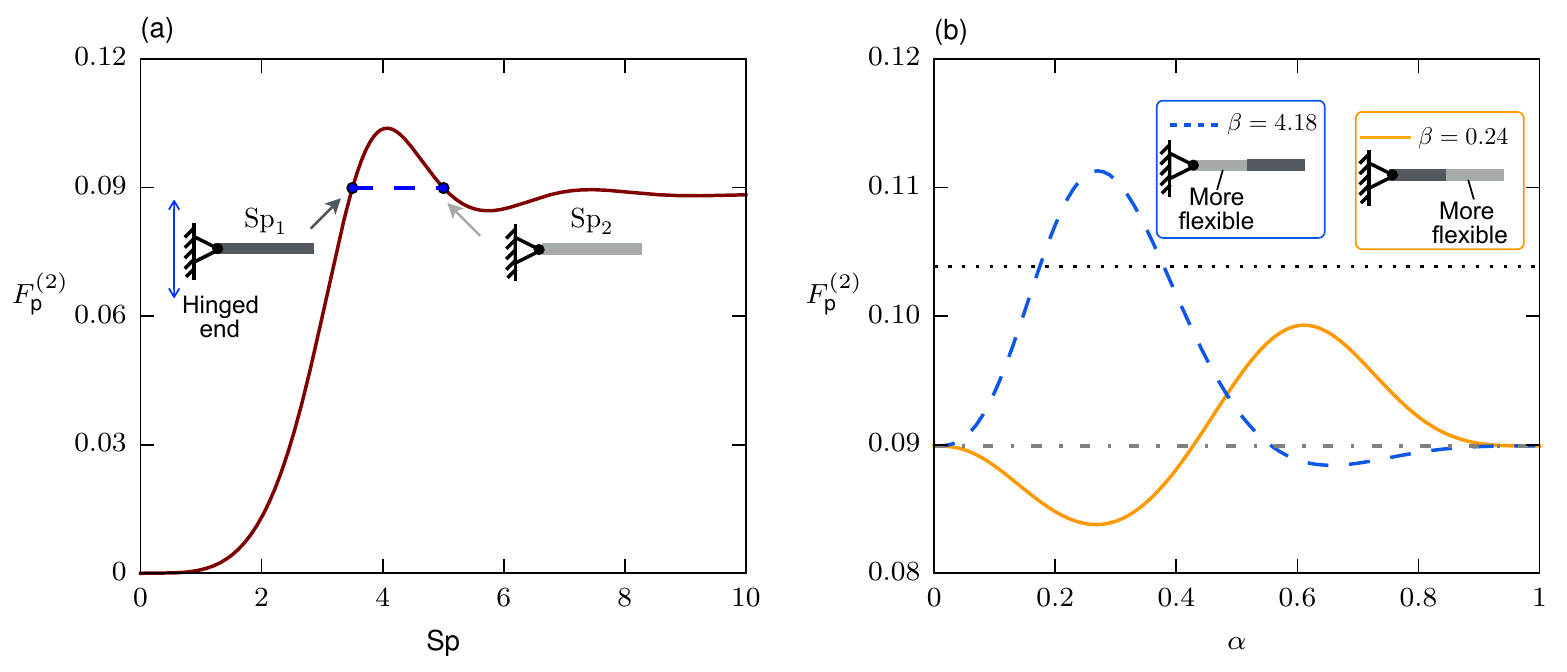}
\caption{\label{fig:elaelaTORQ}(a) Propulsive thrust generated by a filament of uniform bending stiffness under a torque-free displacement actuation \cite{pak2014theoretical}. The two blue dots indicate two filaments with different bending stiffnesses $\text{Sp}=3.5$ and $\text{Sp}=5.006$ generating the same propulsive force.  (b) Propulsive thrust generated by two-segment filaments as a function of $\alpha$ at different values of stiffness ratio $\beta$. For $\beta = 4.18$ ($A_2>A_1$, more flexible material at the actuation end), the maximum propulsive force generated is greater than the maximum achievable thrust of a filament with any uniform stiffness.}
\end{figure*}

For a displacement-driven cantilevered filament, we observed a trend that by incorporating a spatially decaying stiffness one can potentially enhance propulsive performance and this result seems to be consistent with the experimental observations obtained by Maier \textit{et. al.} \cite{Maier2016NANO}, where a flagellar bundle with an exponentially decaying stiffness under a rotational actuation was found to exhibit a larger swimming speed than the uniform stiffness case. However, we shall show in this section that optimal flexibility distribution significantly depends on the boundary condition. Therefore, one cannot simply extend the results by Maier \textit{et. al.} \cite{Maier2016NANO} and this work to other scenarios where the actuation mechanism may differ.

We demonstrate that by only changing the tethering condition from a cantilevered filament to a hinged filament at the actuation end, a qualitatively different trend regarding the optimal flexibility distribution will be obtained. Specifically, we consider a boundary driven passive filament where one end is under torque-free harmonic oscillation while the other end is free \cite{Evans2010PRE}. At the actuation end ($s=0$), we have 
\begin{align}
y(0,t) = \epsilon \sin t,\quad T_\text{ext}(0) = -[A\kappa]_{s=0}=0,
\end{align}
from which we obtain that $\psi_s(0,t)=0$.

Following the cantilevered case, we investigate a two-segment filament undergoing small-amplitude displacement oscillation. The governing equations for $\psi_1^{(1)}$ and $\psi_2^{(1)}$ are the same as those obtained for a cantilevered filament, but with different boundary conditions at the driven end. The leading order propulsive force is given by
\begin{align}
F_\text{p}^{(2)} = &\frac{1-\gamma}{\gamma\text{Sp}_1^4} \biggl< \psi_1^{(1)}(0,t)\partial_s^2\psi_1^{(1)}(0,t)+\frac{1}{2}\left[ (\partial_s\psi_1^{(1)}(\alpha,t))^2 -\beta(\partial_s\psi_2^{(1)}(\alpha,t))^2\right] \biggr>,
\end{align}
which, when $\beta =1$, reduces to the case of uniform bending stiffness along the filament and the propulsive force as a function of $\text{Sp}$ is shown in Fig.~\ref{fig:elaelaTORQ}(a) \cite{pak2014theoretical}. For illustration we consider two bending stiffnesses (corresponding to two different sperm numbers $\text{Sp}=3.5$ and $\text{Sp}=5.006$) that lead to the same propulsive thrust ($F_\text{p}^{(2)}=0.09$, denoted by blue dots in Fig.~\ref{fig:elaelaTORQ}a). With these two bending stiffnesses, two possible configurations of a two-segment filament can be constructed: (i) a more flexible segment at the actuation end ($\beta = 4.18$, dashed line in Fig.~\ref{fig:elaelaTORQ}b) and (ii) a more flexible segment at the free end ($\beta = 0.24$, solid line in Fig.~\ref{fig:elaelaTORQ}b). 

Similar to the cantilevered case, the propulsive thrust varies non-monotonically with the relative portion of the two segments $\alpha$ in both cases, and the propulsive thrust yields the uniform stiffness value ($F_\text{p}^{(2)} = 0.09$) in the limits $\alpha\rightarrow0,1$ (horizontal dash-dotted line in Fig.~\ref{fig:elaelaTORQ}b). However, unlike the case of a cantilevered filament, where putting a more flexible segment at the free end is favorable for propulsion enhancement, for a hinged filament the opposite is true: a more flexible segment at the actuation end ($\beta = 4.18$, dashed line) generates a propulsive thrust greater than the maximum possible thrust with uniform bending stiffness. The other configuration ($\beta = 0.24$, solid line) does not outperform the optimal uniform stiffness case. As an additional example, one can show that an exponentially increasing stiffness can improve the propulsion of a hinged filament compared with the optimal uniform case, in contrary to the case of a cantilevered filament and the rotational actuation case by Maier \textit{et.~al.} \cite{Maier2016NANO}. These qualitatively different behaviors demonstrate the significant dependence of the optimal configuration on the boundary actuation mechanism.

\section{Conclusion}
\label{sec:conclude}
In this paper, we have presented an analytical and numerical investigation on enhancing propulsion of a boundary-driven filament by way of spatially variable flexibility along the filament in the asymptotic limit of small-amplitude oscillations. First, we demonstrated that the torsional spring configuration, which maximizes propulsive thrust in the high Reynolds number regime, does not outperform the optimal uniform stiffness in the low Reynolds number regime. Second, specific continuous and segmented flexibility distributions that generate propulsive thrusts greater than the maximum achievable by any uniform stiffness filament were determined. Third, the specific examples considered in this work suggest that for the case of a cantilevered filament it is favorable to have decreasing stiffness towards the free end, consistent with recent findings regarding enhanced propulsion using a flagellar bundle with an exponentially decreasing stiffness profile under rotational actuation \cite{Maier2016NANO}; however, we emphasize that the favorable flexibility distribution depends significantly on the precise boundary condition at the actuation end. As an example, a hinged filament prefers more flexibility at the actuation end instead. To conclude, we have shown that both the fluid physics (Reynolds number) and mechanical tethering condition can qualitatively modify the optimal flexibility arrangement of a driven filament. Therefore, there is no general design rule for the optimal flexibility distribution, which instead must be determined on a case by case basis.  

The enhancement of propulsion revealed by our study may be useful for the development of synthetic microscopic swimmers requiring large propulsive force for the delivery of therapeutic payloads \cite{pak2011SoftMatter, Gao2012}, penetrating complex media \cite{Nelson2014,Schamel2014} or clearing clogged arteries \cite{nelson2010microrobots,Cheang2015}. A method for fabricating filaments with spatially varying stiffness has already been developed using DNA self-assembly as reported by Maier \textit{et.~al.} \cite{Maier2016NANO}. In addition, multi-segmented hybrid nanostructures may be synthesized by chemical vapor deposition and electrodeposition methods \cite{Shaijumon08}, and these proposed theoretical designs may prompt new experimental implementations. 

As a first step to probe the difference between a filament with non-uniform stiffness versus one with uniform stiffness, the stiffness is varied through modifying the elastic modulus. We note that the flexibility distribution can also change by varying the geometry of the cross-section along the filament. This however complicates the physical picture since any variation in stiffness is then also accompanied by changes in hydrodynamic stresses which may lead to further enhancement (or reduction) of propulsion and is a logical next step for analysis.

\appendix

\section{Numerical approach}
\label{sec:appendix}

\subsection{Numerical solution for continuous stiffness distributions}
\label{subsec:continuous_num}
The ODE for $h(s)$, $i \text{Sp}^{4} h + \left(\Af h_{s}\right)_{sss}=0$,  is a boundary value problem which is solved using Matlab's built-in \texttt{bvp4c} solver for a given stiffness profile. As a validation of the algorithm, the numerical results for both uniform ($\Af= 1$) and two-segment (approximated using hyperbolic tangent, see Sec.~\ref{subsec:two_seg}) stiffness distributions match with the analytical solutions. Once the filament shape is obtained, the propulsive force can be evaluated from Eq.~(\ref{eq:prop_general}) where Gauss-Legendre quadrature is used for numerical integration.

\subsection{Numerical optimization}
\label{subsec:opt}

The optimization over polynomial and exponential stiffness distributions is performed using the \texttt{fmincon} solver in Matlab. For both linear and quadratic distributions, the optimization routine points to a limiting case where the boundary value problem (BVP) for $h(s)$ has a singular point at $s=1$, \textit{i.e.}, the stiffness at the free end goes to zero where the numerical solver itself cannot produce a solution. We then solved these two singular BVPs using the Matlab package \texttt{BVPSUITE} which can handle a singularity of the second kind \cite{kitzhofer2010new,de1976difference,de1980boundary}.

\section{Small amplitude asymptotics}
We present the details of the small-amplitude asymptotic calculations in this appendix. 
Since the magnitude of boundary actuation is $\mathcal{O}(\epsilon)$, we have the tangent angle $\psi \sim \epsilon$. From Eq.~(\ref{eq:govern2}), one then expects the tensile force $\tau\sim \epsilon^2$. We perform regular perturbation expansions $\psi = \psi^{(1)}\epsilon + \psi^{(2)}\epsilon^2+\mathcal{O}(\epsilon^3)$ and $\tau = \tau^{(2)} \epsilon^2 +\mathcal{O}(\epsilon^3) $, and then the leading order equation for $\psi$, upon substitution into Eq.~(\ref{eq:govern1}), is given by $\xi_\perp \psi^{(1)}_t+ \partial_s^3(A\psi^{(1)}_s) = 0$, with the dimensionless version given in Eq.~(\ref{eqn:firstorderdimensionless}).

The force-free  and torque-free conditions at the free end given by Eq.~(\ref{eq:ForceTorqueBC}) translate to $\psi^{(1)}_{ss}(L,t)=0$ and $\psi^{(1)}_{s}(L,t)=0$. At the cantilevered end, we have $\psi^{(1)}(0,t)=0$. The last boundary condition can be obtained by looking at the force balance given by Eq.~(\ref{eq:forceBalance}). Noting that $\bx = y_0\sin\omega t\be_y+\int_0^s \bt(s^\prime,t)\d s^\prime = s\be_x+ \epsilon(\int_0^s\psi^{(1)}\d s + L\sin\omega t)\be_y + \mathcal{O}(\epsilon^2)$, the leading order force balance is at $\be_y$ direction and given by
\begin{align}\label{eq:forcebalance_int}
L\xi_\perp\omega \cos\omega t +\xi_\perp \int_0^s\psi_t^{(1)}\d s+ A\partial_s^3\psi^{(1)}=0.
\end{align}
Evaluating the above equation at $s=0$, we obtain the boundary condition associated with the harmonic actuation, $L\xi_\perp\omega \cos\omega t  +A\partial_s^3\psi^{(1)}(0,t)=0$. We note that governing equation for $\psi^{(1)}$ can also be obtained by differentiating Eq.~(\ref{eq:forcebalance_int}) with respect to arclength. 

For the case of segmented stiffness distribution, similar equations can be obtained by considering the force balance and boundary conditions for each segment following the approach outlined above.

\section*{Acknowledgments}
The authors would like to thank Saverio E. Spagnolie for useful discussions and advice.

\bibliography{REF}

\end{document}

%% file: definitions.tex
\def \d{\text{d}}
\def \diff{\text{d}}

\def \bF{\mathbf{F}}

\def \be{\mathbf{e}}

\def \bf{\mathbf{f}}

\def \bn{\mathbf{n}}

\def \bt{\mathbf{t}}

\def \bx{\mathbf{x}}

\def \Af{\mathcal{A}}

%% file: paper_arxiv.bbl
\begin{thebibliography}{73}%
\makeatletter
\providecommand \@ifxundefined [1]{%
 \@ifx{#1\undefined}
}%
\providecommand \@ifnum [1]{%
 \ifnum #1\expandafter \@firstoftwo
 \else \expandafter \@secondoftwo
 \fi
}%
\providecommand \@ifx [1]{%
 \ifx #1\expandafter \@firstoftwo
 \else \expandafter \@secondoftwo
 \fi
}%
\providecommand \natexlab [1]{#1}%
\providecommand \enquote  [1]{``#1''}%
\providecommand \bibnamefont  [1]{#1}%
\providecommand \bibfnamefont [1]{#1}%
\providecommand \citenamefont [1]{#1}%
\providecommand \href@noop [0]{\@secondoftwo}%
\providecommand \href [0]{\begingroup \@sanitize@url \@href}%
\providecommand \@href[1]{\@@startlink{#1}\@@href}%
\providecommand \@@href[1]{\endgroup#1\@@endlink}%
\providecommand \@sanitize@url [0]{\catcode `\\12\catcode `\$12\catcode
  `\&12\catcode `\#12\catcode `\^12\catcode `\_12\catcode `\%12\relax}%
\providecommand \@@startlink[1]{}%
\providecommand \@@endlink[0]{}%
\providecommand \url  [0]{\begingroup\@sanitize@url \@url }%
\providecommand \@url [1]{\endgroup\@href {#1}{\urlprefix }}%
\providecommand \urlprefix  [0]{URL }%
\providecommand \Eprint [0]{\href }%
\providecommand \doibase [0]{http://dx.doi.org/}%
\providecommand \selectlanguage [0]{\@gobble}%
\providecommand \bibinfo  [0]{\@secondoftwo}%
\providecommand \bibfield  [0]{\@secondoftwo}%
\providecommand \translation [1]{[#1]}%
\providecommand \BibitemOpen [0]{}%
\providecommand \bibitemStop [0]{}%
\providecommand \bibitemNoStop [0]{.\EOS\space}%
\providecommand \EOS [0]{\spacefactor3000\relax}%
\providecommand \BibitemShut  [1]{\csname bibitem#1\endcsname}%
\let\auto@bib@innerbib\@empty
\bibitem [{\citenamefont {Purcell}(1977)}]{purcell1977life}%
  \BibitemOpen
  \bibfield  {author} {\bibinfo {author} {\bibfnamefont {E.~M.}\ \bibnamefont
  {Purcell}},\ }\href {\doibase 10.1119/1.10903} {\bibfield  {journal}
  {\bibinfo  {journal} {Am. J. Phys.}\ }\textbf {\bibinfo {volume} {45}},\
  \bibinfo {pages} {3} (\bibinfo {year} {1977})}\BibitemShut {NoStop}%
\bibitem [{\citenamefont {Brennen}\ and\ \citenamefont
  {Winet}(1977)}]{brennen1977rev}%
  \BibitemOpen
  \bibfield  {author} {\bibinfo {author} {\bibfnamefont {C.}~\bibnamefont
  {Brennen}}\ and\ \bibinfo {author} {\bibfnamefont {H.}~\bibnamefont
  {Winet}},\ }\href {\doibase 10.1146/annurev.fl.09.010177.002011} {\bibfield
  {journal} {\bibinfo  {journal} {Annu. Rev. Fluid Mech.}\ }\textbf {\bibinfo
  {volume} {9}},\ \bibinfo {pages} {339} (\bibinfo {year} {1977})}\BibitemShut
  {NoStop}%
\bibitem [{\citenamefont {Fauci}\ and\ \citenamefont {Dillon}(2006)}]{Fauci06}%
  \BibitemOpen
  \bibfield  {author} {\bibinfo {author} {\bibfnamefont {L.~J.}\ \bibnamefont
  {Fauci}}\ and\ \bibinfo {author} {\bibfnamefont {R.}~\bibnamefont {Dillon}},\
  }\href {\doibase 10.1146/annurev.fluid.37.061903.175725} {\bibfield
  {journal} {\bibinfo  {journal} {Annu. Rev. Fluid Mech.}\ }\textbf {\bibinfo
  {volume} {38}},\ \bibinfo {pages} {371} (\bibinfo {year} {2006})}\BibitemShut
  {NoStop}%
\bibitem [{\citenamefont {Berg}(2008)}]{berg2008coli}%
  \BibitemOpen
  \bibfield  {author} {\bibinfo {author} {\bibfnamefont {H.~C.}\ \bibnamefont
  {Berg}},\ }\href@noop {} {\emph {\bibinfo {title} {E. coli in Motion}}}\
  (\bibinfo  {publisher} {Springer Science \& Business Media},\ \bibinfo {year}
  {2008})\BibitemShut {NoStop}%
\bibitem [{\citenamefont {Lauga}\ and\ \citenamefont
  {Powers}(2009)}]{lauga2009hydrodynamics}%
  \BibitemOpen
  \bibfield  {author} {\bibinfo {author} {\bibfnamefont {E.}~\bibnamefont
  {Lauga}}\ and\ \bibinfo {author} {\bibfnamefont {T.~R.}\ \bibnamefont
  {Powers}},\ }\href {\doibase 10.1088/0034-4885/72/9/096601} {\bibfield
  {journal} {\bibinfo  {journal} {Rep. Prog. Phys.}\ }\textbf {\bibinfo
  {volume} {72}},\ \bibinfo {pages} {096601} (\bibinfo {year}
  {2009})}\BibitemShut {NoStop}%
\bibitem [{\citenamefont {Yadav}\ \emph {et~al.}(2015)\citenamefont {Yadav},
  \citenamefont {Duan}, \citenamefont {Butler},\ and\ \citenamefont
  {Sen}}]{yadav2015anatomy}%
  \BibitemOpen
  \bibfield  {author} {\bibinfo {author} {\bibfnamefont {V.}~\bibnamefont
  {Yadav}}, \bibinfo {author} {\bibfnamefont {W.}~\bibnamefont {Duan}},
  \bibinfo {author} {\bibfnamefont {P.~J.}\ \bibnamefont {Butler}}, \ and\
  \bibinfo {author} {\bibfnamefont {A.}~\bibnamefont {Sen}},\ }\href {\doibase
  10.1146/annurev-biophys-060414-034216} {\bibfield  {journal} {\bibinfo
  {journal} {Annu. Rev. Biophys.}\ }\textbf {\bibinfo {volume} {44}},\ \bibinfo
  {pages} {77} (\bibinfo {year} {2015})}\BibitemShut {NoStop}%
\bibitem [{\citenamefont {Elgeti}\ \emph {et~al.}(2015)\citenamefont {Elgeti},
  \citenamefont {Winkler},\ and\ \citenamefont {Gompper}}]{Elgeti2015}%
  \BibitemOpen
  \bibfield  {author} {\bibinfo {author} {\bibfnamefont {J.}~\bibnamefont
  {Elgeti}}, \bibinfo {author} {\bibfnamefont {R.~G.}\ \bibnamefont {Winkler}},
  \ and\ \bibinfo {author} {\bibfnamefont {G.}~\bibnamefont {Gompper}},\ }\href
  {\doibase 10.1088/0034-4885/78/5/056601} {\bibfield  {journal} {\bibinfo
  {journal} {Rep. Prog. Phys.}\ }\textbf {\bibinfo {volume} {78}},\ \bibinfo
  {pages} {056601} (\bibinfo {year} {2015})}\BibitemShut {NoStop}%
\bibitem [{\citenamefont {Dreyfus}\ \emph {et~al.}(2005)\citenamefont
  {Dreyfus}, \citenamefont {Baudry}, \citenamefont {Roper}, \citenamefont
  {Fermigier}, \citenamefont {Stone},\ and\ \citenamefont
  {Bibette}}]{dreyfus2005microscopic}%
  \BibitemOpen
  \bibfield  {author} {\bibinfo {author} {\bibfnamefont {R.}~\bibnamefont
  {Dreyfus}}, \bibinfo {author} {\bibfnamefont {J.}~\bibnamefont {Baudry}},
  \bibinfo {author} {\bibfnamefont {M.~L.}\ \bibnamefont {Roper}}, \bibinfo
  {author} {\bibfnamefont {M.}~\bibnamefont {Fermigier}}, \bibinfo {author}
  {\bibfnamefont {H.~A.}\ \bibnamefont {Stone}}, \ and\ \bibinfo {author}
  {\bibfnamefont {J.}~\bibnamefont {Bibette}},\ }\href {\doibase
  10.1038/nature04090} {\bibfield  {journal} {\bibinfo  {journal} {Nature}\
  }\textbf {\bibinfo {volume} {437}},\ \bibinfo {pages} {862} (\bibinfo {year}
  {2005})}\BibitemShut {NoStop}%
\bibitem [{\citenamefont {Pak}\ \emph {et~al.}(2011)\citenamefont {Pak},
  \citenamefont {Gao}, \citenamefont {Wang},\ and\ \citenamefont
  {Lauga}}]{pak2011SoftMatter}%
  \BibitemOpen
  \bibfield  {author} {\bibinfo {author} {\bibfnamefont {O.~S.}\ \bibnamefont
  {Pak}}, \bibinfo {author} {\bibfnamefont {W.}~\bibnamefont {Gao}}, \bibinfo
  {author} {\bibfnamefont {J.}~\bibnamefont {Wang}}, \ and\ \bibinfo {author}
  {\bibfnamefont {E.}~\bibnamefont {Lauga}},\ }\href {\doibase
  10.1039/C1SM05503H} {\bibfield  {journal} {\bibinfo  {journal} {Soft Matter}\
  }\textbf {\bibinfo {volume} {7}},\ \bibinfo {pages} {8169} (\bibinfo {year}
  {2011})}\BibitemShut {NoStop}%
\bibitem [{\citenamefont {Williams}\ \emph {et~al.}(2014)\citenamefont
  {Williams}, \citenamefont {Anand}, \citenamefont {Rajagopalan},\ and\
  \citenamefont {Saif}}]{williams2014Nature}%
  \BibitemOpen
  \bibfield  {author} {\bibinfo {author} {\bibfnamefont {B.~J.}\ \bibnamefont
  {Williams}}, \bibinfo {author} {\bibfnamefont {S.~V.}\ \bibnamefont {Anand}},
  \bibinfo {author} {\bibfnamefont {J.}~\bibnamefont {Rajagopalan}}, \ and\
  \bibinfo {author} {\bibfnamefont {M.~T.~A.}\ \bibnamefont {Saif}},\ }\href
  {\doibase 10.1038/ncomms4081} {\bibfield  {journal} {\bibinfo  {journal}
  {Nat. Commun.}\ }\textbf {\bibinfo {volume} {5}},\ \bibinfo {pages} {3081}
  (\bibinfo {year} {2014})}\BibitemShut {NoStop}%
\bibitem [{\citenamefont {Maier}\ \emph {et~al.}(2016)\citenamefont {Maier},
  \citenamefont {Weig}, \citenamefont {Oswald}, \citenamefont {Frey},
  \citenamefont {Fischer},\ and\ \citenamefont {Liedl}}]{Maier2016NANO}%
  \BibitemOpen
  \bibfield  {author} {\bibinfo {author} {\bibfnamefont {A.~M.}\ \bibnamefont
  {Maier}}, \bibinfo {author} {\bibfnamefont {C.}~\bibnamefont {Weig}},
  \bibinfo {author} {\bibfnamefont {P.}~\bibnamefont {Oswald}}, \bibinfo
  {author} {\bibfnamefont {E.}~\bibnamefont {Frey}}, \bibinfo {author}
  {\bibfnamefont {P.}~\bibnamefont {Fischer}}, \ and\ \bibinfo {author}
  {\bibfnamefont {T.}~\bibnamefont {Liedl}},\ }\href {\doibase
  10.1021/acs.nanolett.5b03716} {\bibfield  {journal} {\bibinfo  {journal}
  {Nano Lett.}\ }\textbf {\bibinfo {volume} {16}},\ \bibinfo {pages} {906}
  (\bibinfo {year} {2016})}\BibitemShut {NoStop}%
\bibitem [{\citenamefont {Ebbens}\ and\ \citenamefont
  {Howse}(2010)}]{Ebbens2010}%
  \BibitemOpen
  \bibfield  {author} {\bibinfo {author} {\bibfnamefont {S.~J.}\ \bibnamefont
  {Ebbens}}\ and\ \bibinfo {author} {\bibfnamefont {J.~R.}\ \bibnamefont
  {Howse}},\ }\href {\doibase 10.1039/B918598D} {\bibfield  {journal} {\bibinfo
   {journal} {Soft Matter}\ }\textbf {\bibinfo {volume} {6}},\ \bibinfo {pages}
  {726} (\bibinfo {year} {2010})}\BibitemShut {NoStop}%
\bibitem [{\citenamefont {Nelson}\ \emph {et~al.}(2010)\citenamefont {Nelson},
  \citenamefont {Kaliakatsos},\ and\ \citenamefont
  {Abbott}}]{nelson2010microrobots}%
  \BibitemOpen
  \bibfield  {author} {\bibinfo {author} {\bibfnamefont {B.~J.}\ \bibnamefont
  {Nelson}}, \bibinfo {author} {\bibfnamefont {I.~K.}\ \bibnamefont
  {Kaliakatsos}}, \ and\ \bibinfo {author} {\bibfnamefont {J.~J.}\ \bibnamefont
  {Abbott}},\ }\href {\doibase 10.1146/annurev-bioeng-010510-103409} {\bibfield
   {journal} {\bibinfo  {journal} {Annu. Rev. Biomed. Eng.}\ }\textbf {\bibinfo
  {volume} {12}},\ \bibinfo {pages} {55} (\bibinfo {year} {2010})}\BibitemShut
  {NoStop}%
\bibitem [{\citenamefont {Nelson}\ and\ \citenamefont
  {Peyer}(2014)}]{Nelson2014}%
  \BibitemOpen
  \bibfield  {author} {\bibinfo {author} {\bibfnamefont {B.~J.}\ \bibnamefont
  {Nelson}}\ and\ \bibinfo {author} {\bibfnamefont {K.~E.}\ \bibnamefont
  {Peyer}},\ }\href {\doibase 10.1021/nn504295z} {\bibfield  {journal}
  {\bibinfo  {journal} {ACS Nano}\ }\textbf {\bibinfo {volume} {8}},\ \bibinfo
  {pages} {8718} (\bibinfo {year} {2014})}\BibitemShut {NoStop}%
\bibitem [{\citenamefont {Tierno}\ \emph {et~al.}(2008)\citenamefont {Tierno},
  \citenamefont {Golestanian}, \citenamefont {Pagonabarraga},\ and\
  \citenamefont {Sagu\'es}}]{Tierno2008}%
  \BibitemOpen
  \bibfield  {author} {\bibinfo {author} {\bibfnamefont {P.}~\bibnamefont
  {Tierno}}, \bibinfo {author} {\bibfnamefont {R.}~\bibnamefont {Golestanian}},
  \bibinfo {author} {\bibfnamefont {I.}~\bibnamefont {Pagonabarraga}}, \ and\
  \bibinfo {author} {\bibfnamefont {F.}~\bibnamefont {Sagu\'es}},\ }\href
  {\doibase 10.1103/PhysRevLett.101.218304} {\bibfield  {journal} {\bibinfo
  {journal} {Phys. Rev. Lett.}\ }\textbf {\bibinfo {volume} {101}},\ \bibinfo
  {pages} {218304} (\bibinfo {year} {2008})}\BibitemShut {NoStop}%
\bibitem [{\citenamefont {Sing}\ \emph {et~al.}(2010)\citenamefont {Sing},
  \citenamefont {Schmid}, \citenamefont {Schneider}, \citenamefont {Franke},\
  and\ \citenamefont {Alexander-Katz}}]{Sing2010}%
  \BibitemOpen
  \bibfield  {author} {\bibinfo {author} {\bibfnamefont {C.~E.}\ \bibnamefont
  {Sing}}, \bibinfo {author} {\bibfnamefont {L.}~\bibnamefont {Schmid}},
  \bibinfo {author} {\bibfnamefont {M.~F.}\ \bibnamefont {Schneider}}, \bibinfo
  {author} {\bibfnamefont {T.}~\bibnamefont {Franke}}, \ and\ \bibinfo {author}
  {\bibfnamefont {A.}~\bibnamefont {Alexander-Katz}},\ }\href {\doibase
  10.1073/pnas.0906489107} {\bibfield  {journal} {\bibinfo  {journal} {Proc.
  Natl. Acad. Sci. U. S. A.}\ }\textbf {\bibinfo {volume} {107}},\ \bibinfo
  {pages} {535} (\bibinfo {year} {2010})}\BibitemShut {NoStop}%
\bibitem [{\citenamefont {Zhang}\ \emph {et~al.}(2010)\citenamefont {Zhang},
  \citenamefont {Petit}, \citenamefont {Lu}, \citenamefont {Kratochvil},
  \citenamefont {Peyer}, \citenamefont {Pei}, \citenamefont {Lou},\ and\
  \citenamefont {Nelson}}]{Zhang2010}%
  \BibitemOpen
  \bibfield  {author} {\bibinfo {author} {\bibfnamefont {L.}~\bibnamefont
  {Zhang}}, \bibinfo {author} {\bibfnamefont {T.}~\bibnamefont {Petit}},
  \bibinfo {author} {\bibfnamefont {Y.}~\bibnamefont {Lu}}, \bibinfo {author}
  {\bibfnamefont {B.~E.}\ \bibnamefont {Kratochvil}}, \bibinfo {author}
  {\bibfnamefont {K.~E.}\ \bibnamefont {Peyer}}, \bibinfo {author}
  {\bibfnamefont {R.}~\bibnamefont {Pei}}, \bibinfo {author} {\bibfnamefont
  {J.}~\bibnamefont {Lou}}, \ and\ \bibinfo {author} {\bibfnamefont {B.~J.}\
  \bibnamefont {Nelson}},\ }\href {\doibase 10.1021/nn101861n} {\bibfield
  {journal} {\bibinfo  {journal} {ACS Nano}\ }\textbf {\bibinfo {volume} {4}},\
  \bibinfo {pages} {6228} (\bibinfo {year} {2010})}\BibitemShut {NoStop}%
\bibitem [{\citenamefont {Zhu}\ \emph {et~al.}(2013)\citenamefont {Zhu},
  \citenamefont {Lauga},\ and\ \citenamefont {Brandt}}]{zhu2013}%
  \BibitemOpen
  \bibfield  {author} {\bibinfo {author} {\bibfnamefont {L.}~\bibnamefont
  {Zhu}}, \bibinfo {author} {\bibfnamefont {E.}~\bibnamefont {Lauga}}, \ and\
  \bibinfo {author} {\bibfnamefont {L.}~\bibnamefont {Brandt}},\ }\href
  {\doibase 10.1017/jfm.2013.225} {\bibfield  {journal} {\bibinfo  {journal}
  {J. Fluid Mech.}\ }\textbf {\bibinfo {volume} {726}},\ \bibinfo {pages} {285}
  (\bibinfo {year} {2013})}\BibitemShut {NoStop}%
\bibitem [{\citenamefont {Ledesma-Aguilar}\ and\ \citenamefont
  {Yeomans}(2013)}]{Aguilar2013}%
  \BibitemOpen
  \bibfield  {author} {\bibinfo {author} {\bibfnamefont {R.}~\bibnamefont
  {Ledesma-Aguilar}}\ and\ \bibinfo {author} {\bibfnamefont {J.~M.}\
  \bibnamefont {Yeomans}},\ }\href {\doibase 10.1103/PhysRevLett.111.138101}
  {\bibfield  {journal} {\bibinfo  {journal} {Phys. Rev. Lett.}\ }\textbf
  {\bibinfo {volume} {111}},\ \bibinfo {pages} {138101} (\bibinfo {year}
  {2013})}\BibitemShut {NoStop}%
\bibitem [{\citenamefont {Leshansky}(2009)}]{Leshansky2009}%
  \BibitemOpen
  \bibfield  {author} {\bibinfo {author} {\bibfnamefont {A.~M.}\ \bibnamefont
  {Leshansky}},\ }\href {\doibase 10.1103/PhysRevE.80.051911} {\bibfield
  {journal} {\bibinfo  {journal} {Phys. Rev. E}\ }\textbf {\bibinfo {volume}
  {80}},\ \bibinfo {pages} {051911} (\bibinfo {year} {2009})}\BibitemShut
  {NoStop}%
\bibitem [{\citenamefont {Fu}\ \emph {et~al.}(2010)\citenamefont {Fu},
  \citenamefont {Shenoy},\ and\ \citenamefont {Powers}}]{Fu2010}%
  \BibitemOpen
  \bibfield  {author} {\bibinfo {author} {\bibfnamefont {H.~C.}\ \bibnamefont
  {Fu}}, \bibinfo {author} {\bibfnamefont {V.~B.}\ \bibnamefont {Shenoy}}, \
  and\ \bibinfo {author} {\bibfnamefont {T.~R.}\ \bibnamefont {Powers}},\
  }\href {\doibase 10.1209/0295-5075/91/24002} {\bibfield  {journal} {\bibinfo
  {journal} {Europhys. Lett.}\ }\textbf {\bibinfo {volume} {91}},\ \bibinfo
  {pages} {24002} (\bibinfo {year} {2010})}\BibitemShut {NoStop}%
\bibitem [{\citenamefont {Teran}\ \emph {et~al.}(2010)\citenamefont {Teran},
  \citenamefont {Fauci},\ and\ \citenamefont {Shelley}}]{Teran2010}%
  \BibitemOpen
  \bibfield  {author} {\bibinfo {author} {\bibfnamefont {J.}~\bibnamefont
  {Teran}}, \bibinfo {author} {\bibfnamefont {L.}~\bibnamefont {Fauci}}, \ and\
  \bibinfo {author} {\bibfnamefont {M.}~\bibnamefont {Shelley}},\ }\href
  {\doibase 10.1103/PhysRevLett.104.038101} {\bibfield  {journal} {\bibinfo
  {journal} {Phys. Rev. Lett.}\ }\textbf {\bibinfo {volume} {104}},\ \bibinfo
  {pages} {038101} (\bibinfo {year} {2010})}\BibitemShut {NoStop}%
\bibitem [{\citenamefont {Liu}\ \emph {et~al.}(2011)\citenamefont {Liu},
  \citenamefont {Powers},\ and\ \citenamefont {Breuer}}]{Liu2011}%
  \BibitemOpen
  \bibfield  {author} {\bibinfo {author} {\bibfnamefont {B.}~\bibnamefont
  {Liu}}, \bibinfo {author} {\bibfnamefont {T.~R.}\ \bibnamefont {Powers}}, \
  and\ \bibinfo {author} {\bibfnamefont {K.~S.}\ \bibnamefont {Breuer}},\
  }\href {\doibase 10.1073/pnas.1113082108} {\bibfield  {journal} {\bibinfo
  {journal} {Proc. Natl. Acad. Sci. U. S. A.}\ }\textbf {\bibinfo {volume}
  {108}},\ \bibinfo {pages} {19516} (\bibinfo {year} {2011})}\BibitemShut
  {NoStop}%
\bibitem [{\citenamefont {Keim}\ \emph {et~al.}(2012)\citenamefont {Keim},
  \citenamefont {Garcia},\ and\ \citenamefont {Arratia}}]{Nathan2012}%
  \BibitemOpen
  \bibfield  {author} {\bibinfo {author} {\bibfnamefont {N.~C.}\ \bibnamefont
  {Keim}}, \bibinfo {author} {\bibfnamefont {M.}~\bibnamefont {Garcia}}, \ and\
  \bibinfo {author} {\bibfnamefont {P.~E.}\ \bibnamefont {Arratia}},\ }\href
  {\doibase 10.1063/1.4746792} {\bibfield  {journal} {\bibinfo  {journal}
  {Phys. Fluids}\ }\textbf {\bibinfo {volume} {24}},\ \bibinfo {pages} {081703}
  (\bibinfo {year} {2012})}\BibitemShut {NoStop}%
\bibitem [{\citenamefont {Espinosa-Garcia}\ \emph {et~al.}(2013)\citenamefont
  {Espinosa-Garcia}, \citenamefont {Lauga},\ and\ \citenamefont
  {Zenit}}]{Garcia2013}%
  \BibitemOpen
  \bibfield  {author} {\bibinfo {author} {\bibfnamefont {J.}~\bibnamefont
  {Espinosa-Garcia}}, \bibinfo {author} {\bibfnamefont {E.}~\bibnamefont
  {Lauga}}, \ and\ \bibinfo {author} {\bibfnamefont {R.}~\bibnamefont
  {Zenit}},\ }\href {\doibase 10.1063/1.4795166} {\bibfield  {journal}
  {\bibinfo  {journal} {Phys. Fluids}\ }\textbf {\bibinfo {volume} {25}},\
  \bibinfo {pages} {031701} (\bibinfo {year} {2013})}\BibitemShut {NoStop}%
\bibitem [{\citenamefont {Spagnolie}\ \emph {et~al.}(2013)\citenamefont
  {Spagnolie}, \citenamefont {Liu},\ and\ \citenamefont
  {Powers}}]{Spagnolie2013}%
  \BibitemOpen
  \bibfield  {author} {\bibinfo {author} {\bibfnamefont {S.~E.}\ \bibnamefont
  {Spagnolie}}, \bibinfo {author} {\bibfnamefont {B.}~\bibnamefont {Liu}}, \
  and\ \bibinfo {author} {\bibfnamefont {T.~R.}\ \bibnamefont {Powers}},\
  }\href {\doibase 10.1103/PhysRevLett.111.068101} {\bibfield  {journal}
  {\bibinfo  {journal} {Phys. Rev. Lett.}\ }\textbf {\bibinfo {volume} {111}},\
  \bibinfo {pages} {068101} (\bibinfo {year} {2013})}\BibitemShut {NoStop}%
\bibitem [{\citenamefont {Riley}\ and\ \citenamefont
  {Lauga}(2014)}]{Riley2014}%
  \BibitemOpen
  \bibfield  {author} {\bibinfo {author} {\bibfnamefont {E.~E.}\ \bibnamefont
  {Riley}}\ and\ \bibinfo {author} {\bibfnamefont {E.}~\bibnamefont {Lauga}},\
  }\href {\doibase 10.1209/0295-5075/108/34003} {\bibfield  {journal} {\bibinfo
   {journal} {Europhys. Lett.}\ }\textbf {\bibinfo {volume} {108}},\ \bibinfo
  {pages} {34003} (\bibinfo {year} {2014})}\BibitemShut {NoStop}%
\bibitem [{\citenamefont {Thomases}\ and\ \citenamefont
  {Guy}(2014)}]{Thomases2014}%
  \BibitemOpen
  \bibfield  {author} {\bibinfo {author} {\bibfnamefont {B.}~\bibnamefont
  {Thomases}}\ and\ \bibinfo {author} {\bibfnamefont {R.~D.}\ \bibnamefont
  {Guy}},\ }\href {\doibase 10.1103/PhysRevLett.113.098102} {\bibfield
  {journal} {\bibinfo  {journal} {Phys. Rev. Lett.}\ }\textbf {\bibinfo
  {volume} {113}},\ \bibinfo {pages} {098102} (\bibinfo {year}
  {2014})}\BibitemShut {NoStop}%
\bibitem [{\citenamefont {Wr{\' o}bel}\ \emph {et~al.}(2016)\citenamefont
  {Wr{\' o}bel}, \citenamefont {Lynch}, \citenamefont {Barrett}, \citenamefont
  {Fauci},\ and\ \citenamefont {Cortez}}]{wrbel2016}%
  \BibitemOpen
  \bibfield  {author} {\bibinfo {author} {\bibfnamefont {J.~K.}\ \bibnamefont
  {Wr{\' o}bel}}, \bibinfo {author} {\bibfnamefont {S.}~\bibnamefont {Lynch}},
  \bibinfo {author} {\bibfnamefont {A.}~\bibnamefont {Barrett}}, \bibinfo
  {author} {\bibfnamefont {L.}~\bibnamefont {Fauci}}, \ and\ \bibinfo {author}
  {\bibfnamefont {R.}~\bibnamefont {Cortez}},\ }\href {\doibase
  10.1017/jfm.2016.99} {\bibfield  {journal} {\bibinfo  {journal} {J. Fluid
  Mech.}\ }\textbf {\bibinfo {volume} {792}},\ \bibinfo {pages} {775} (\bibinfo
  {year} {2016})}\BibitemShut {NoStop}%
\bibitem [{\citenamefont {Montenegro-Johnson}\ \emph
  {et~al.}(2013)\citenamefont {Montenegro-Johnson}, \citenamefont {Smith},\
  and\ \citenamefont {Loghin}}]{Johnson2013}%
  \BibitemOpen
  \bibfield  {author} {\bibinfo {author} {\bibfnamefont {T.~D.}\ \bibnamefont
  {Montenegro-Johnson}}, \bibinfo {author} {\bibfnamefont {D.~J.}\ \bibnamefont
  {Smith}}, \ and\ \bibinfo {author} {\bibfnamefont {D.}~\bibnamefont
  {Loghin}},\ }\href {\doibase 10.1063/1.4818640} {\bibfield  {journal}
  {\bibinfo  {journal} {Phys. Fluids}\ }\textbf {\bibinfo {volume} {25}},\
  \bibinfo {pages} {081903} (\bibinfo {year} {2013})}\BibitemShut {NoStop}%
\bibitem [{\citenamefont {Qiu}\ \emph {et~al.}(2014)\citenamefont {Qiu},
  \citenamefont {Lee}, \citenamefont {Mark}, \citenamefont {Morozov},
  \citenamefont {M{\"u}nster}, \citenamefont {Mierka}, \citenamefont {Turek},
  \citenamefont {Leshansky},\ and\ \citenamefont {Fischer}}]{qiu2014swimming}%
  \BibitemOpen
  \bibfield  {author} {\bibinfo {author} {\bibfnamefont {T.}~\bibnamefont
  {Qiu}}, \bibinfo {author} {\bibfnamefont {T.-C.}\ \bibnamefont {Lee}},
  \bibinfo {author} {\bibfnamefont {A.~G.}\ \bibnamefont {Mark}}, \bibinfo
  {author} {\bibfnamefont {K.~I.}\ \bibnamefont {Morozov}}, \bibinfo {author}
  {\bibfnamefont {R.}~\bibnamefont {M{\"u}nster}}, \bibinfo {author}
  {\bibfnamefont {O.}~\bibnamefont {Mierka}}, \bibinfo {author} {\bibfnamefont
  {S.}~\bibnamefont {Turek}}, \bibinfo {author} {\bibfnamefont {A.~M.}\
  \bibnamefont {Leshansky}}, \ and\ \bibinfo {author} {\bibfnamefont
  {P.}~\bibnamefont {Fischer}},\ }\href {\doibase 10.1038/ncomms6119}
  {\bibfield  {journal} {\bibinfo  {journal} {Nat. Commun.}\ }\textbf {\bibinfo
  {volume} {5}},\ \bibinfo {pages} {5119} (\bibinfo {year} {2014})}\BibitemShut
  {NoStop}%
\bibitem [{\citenamefont {Li}\ and\ \citenamefont {Ardekani}(2015)}]{Li2015}%
  \BibitemOpen
  \bibfield  {author} {\bibinfo {author} {\bibfnamefont {G.}~\bibnamefont
  {Li}}\ and\ \bibinfo {author} {\bibfnamefont {A.~M.}\ \bibnamefont
  {Ardekani}},\ }\href {\doibase 10.1017/jfm.2015.595} {\bibfield  {journal}
  {\bibinfo  {journal} {J. Fluid Mech.}\ }\textbf {\bibinfo {volume} {784}},\
  \bibinfo {pages} {R4} (\bibinfo {year} {2015})}\BibitemShut {NoStop}%
\bibitem [{\citenamefont {Datt}\ \emph {et~al.}(2015)\citenamefont {Datt},
  \citenamefont {Zhu}, \citenamefont {Elfring},\ and\ \citenamefont
  {Pak}}]{Datt2015}%
  \BibitemOpen
  \bibfield  {author} {\bibinfo {author} {\bibfnamefont {C.}~\bibnamefont
  {Datt}}, \bibinfo {author} {\bibfnamefont {L.}~\bibnamefont {Zhu}}, \bibinfo
  {author} {\bibfnamefont {G.~J.}\ \bibnamefont {Elfring}}, \ and\ \bibinfo
  {author} {\bibfnamefont {O.~S.}\ \bibnamefont {Pak}},\ }\href {\doibase
  10.1017/jfm.2015.600} {\bibfield  {journal} {\bibinfo  {journal} {J. Fluid
  Mech.}\ }\textbf {\bibinfo {volume} {784}},\ \bibinfo {pages} {R1} (\bibinfo
  {year} {2015})}\BibitemShut {NoStop}%
\bibitem [{\citenamefont {Park}\ \emph {et~al.}(2016)\citenamefont {Park},
  \citenamefont {Kim}, \citenamefont {Shin},\ and\ \citenamefont
  {Weitz}}]{Park2016}%
  \BibitemOpen
  \bibfield  {author} {\bibinfo {author} {\bibfnamefont {J.-S.}\ \bibnamefont
  {Park}}, \bibinfo {author} {\bibfnamefont {D.}~\bibnamefont {Kim}}, \bibinfo
  {author} {\bibfnamefont {J.~H.}\ \bibnamefont {Shin}}, \ and\ \bibinfo
  {author} {\bibfnamefont {D.~A.}\ \bibnamefont {Weitz}},\ }\href {\doibase
  10.1039/C5SM01824B} {\bibfield  {journal} {\bibinfo  {journal} {Soft Matter}\
  }\textbf {\bibinfo {volume} {12}},\ \bibinfo {pages} {1892} (\bibinfo {year}
  {2016})}\BibitemShut {NoStop}%
\bibitem [{\citenamefont {Machin}(1958)}]{Machin1958}%
  \BibitemOpen
  \bibfield  {author} {\bibinfo {author} {\bibfnamefont {K.~E.}\ \bibnamefont
  {Machin}},\ }\href@noop {} {\bibfield  {journal} {\bibinfo  {journal} {J.
  Exp. Biol.}\ }\textbf {\bibinfo {volume} {35}},\ \bibinfo {pages} {796}
  (\bibinfo {year} {1958})}\BibitemShut {NoStop}%
\bibitem [{\citenamefont {Wiggins}\ \emph {et~al.}(1998)\citenamefont
  {Wiggins}, \citenamefont {Riveline}, \citenamefont {Ott},\ and\ \citenamefont
  {Goldstein}}]{Wiggins1998}%
  \BibitemOpen
  \bibfield  {author} {\bibinfo {author} {\bibfnamefont {C.~H.}\ \bibnamefont
  {Wiggins}}, \bibinfo {author} {\bibfnamefont {D.}~\bibnamefont {Riveline}},
  \bibinfo {author} {\bibfnamefont {A.}~\bibnamefont {Ott}}, \ and\ \bibinfo
  {author} {\bibfnamefont {R.~E.}\ \bibnamefont {Goldstein}},\ }\href {\doibase
  10.1016/S0006-3495(98)74029-9} {\bibfield  {journal} {\bibinfo  {journal}
  {Biophys. J.}\ }\textbf {\bibinfo {volume} {74}},\ \bibinfo {pages} {1043 }
  (\bibinfo {year} {1998})}\BibitemShut {NoStop}%
\bibitem [{\citenamefont {Wiggins}\ and\ \citenamefont
  {Goldstein}(1998)}]{WigginsPRL1998}%
  \BibitemOpen
  \bibfield  {author} {\bibinfo {author} {\bibfnamefont {C.~H.}\ \bibnamefont
  {Wiggins}}\ and\ \bibinfo {author} {\bibfnamefont {R.~E.}\ \bibnamefont
  {Goldstein}},\ }\href {\doibase 10.1103/PhysRevLett.80.3879} {\bibfield
  {journal} {\bibinfo  {journal} {Phys. Rev. Lett.}\ }\textbf {\bibinfo
  {volume} {80}},\ \bibinfo {pages} {3879} (\bibinfo {year}
  {1998})}\BibitemShut {NoStop}%
\bibitem [{\citenamefont {Lowe}(2003)}]{Lowe2003}%
  \BibitemOpen
  \bibfield  {author} {\bibinfo {author} {\bibfnamefont {C.~P.}\ \bibnamefont
  {Lowe}},\ }\href {\doibase 10.1098/rstb.2003.1340} {\bibfield  {journal}
  {\bibinfo  {journal} {Phil. Trans. R. Soc. B}\ }\textbf {\bibinfo {volume}
  {358}},\ \bibinfo {pages} {1543} (\bibinfo {year} {2003})}\BibitemShut
  {NoStop}%
\bibitem [{\citenamefont {Manghi}\ \emph {et~al.}(2006)\citenamefont {Manghi},
  \citenamefont {Schlagberger},\ and\ \citenamefont {Netz}}]{Manghi2006}%
  \BibitemOpen
  \bibfield  {author} {\bibinfo {author} {\bibfnamefont {M.}~\bibnamefont
  {Manghi}}, \bibinfo {author} {\bibfnamefont {X.}~\bibnamefont
  {Schlagberger}}, \ and\ \bibinfo {author} {\bibfnamefont {R.~R.}\
  \bibnamefont {Netz}},\ }\href {\doibase 10.1103/PhysRevLett.96.068101}
  {\bibfield  {journal} {\bibinfo  {journal} {Phys. Rev. Lett.}\ }\textbf
  {\bibinfo {volume} {96}},\ \bibinfo {pages} {068101} (\bibinfo {year}
  {2006})}\BibitemShut {NoStop}%
\bibitem [{\citenamefont {Gauger}\ and\ \citenamefont
  {Stark}(2006)}]{Gauger2006}%
  \BibitemOpen
  \bibfield  {author} {\bibinfo {author} {\bibfnamefont {E.}~\bibnamefont
  {Gauger}}\ and\ \bibinfo {author} {\bibfnamefont {H.}~\bibnamefont {Stark}},\
  }\href {\doibase 10.1103/PhysRevE.74.021907} {\bibfield  {journal} {\bibinfo
  {journal} {Phys. Rev. E}\ }\textbf {\bibinfo {volume} {74}},\ \bibinfo
  {pages} {021907} (\bibinfo {year} {2006})}\BibitemShut {NoStop}%
\bibitem [{\citenamefont {Lauga}(2007)}]{Lauga2007PRE}%
  \BibitemOpen
  \bibfield  {author} {\bibinfo {author} {\bibfnamefont {E.}~\bibnamefont
  {Lauga}},\ }\href {\doibase 10.1103/PhysRevE.75.041916} {\bibfield  {journal}
  {\bibinfo  {journal} {Phys. Rev. E}\ }\textbf {\bibinfo {volume} {75}},\
  \bibinfo {pages} {041916} (\bibinfo {year} {2007})}\BibitemShut {NoStop}%
\bibitem [{\citenamefont {Coq}\ \emph {et~al.}(2008)\citenamefont {Coq},
  \citenamefont {du~Roure}, \citenamefont {Marthelot}, \citenamefont
  {Bartolo},\ and\ \citenamefont {Fermigier}}]{Coq2008}%
  \BibitemOpen
  \bibfield  {author} {\bibinfo {author} {\bibfnamefont {N.}~\bibnamefont
  {Coq}}, \bibinfo {author} {\bibfnamefont {O.}~\bibnamefont {du~Roure}},
  \bibinfo {author} {\bibfnamefont {J.}~\bibnamefont {Marthelot}}, \bibinfo
  {author} {\bibfnamefont {D.}~\bibnamefont {Bartolo}}, \ and\ \bibinfo
  {author} {\bibfnamefont {M.}~\bibnamefont {Fermigier}},\ }\href {\doibase
  http://dx.doi.org/10.1063/1.2909603} {\bibfield  {journal} {\bibinfo
  {journal} {Phys. Fluids}\ }\textbf {\bibinfo {volume} {20}},\ \bibinfo
  {pages} {051703} (\bibinfo {year} {2008})}\BibitemShut {NoStop}%
\bibitem [{\citenamefont {Keaveny}\ and\ \citenamefont
  {Maxey}(2008)}]{Keaveny08}%
  \BibitemOpen
  \bibfield  {author} {\bibinfo {author} {\bibfnamefont {E.~E.}\ \bibnamefont
  {Keaveny}}\ and\ \bibinfo {author} {\bibfnamefont {M.~R.}\ \bibnamefont
  {Maxey}},\ }\href {\doibase 10.1017/S0022112007009949} {\bibfield  {journal}
  {\bibinfo  {journal} {J. Fluid Mech.}\ }\textbf {\bibinfo {volume} {598}},\
  \bibinfo {pages} {293} (\bibinfo {year} {2008})}\BibitemShut {NoStop}%
\bibitem [{\citenamefont {Fu}\ \emph {et~al.}(2008)\citenamefont {Fu},
  \citenamefont {Wolgemuth},\ and\ \citenamefont {Powers}}]{Fu2008}%
  \BibitemOpen
  \bibfield  {author} {\bibinfo {author} {\bibfnamefont {H.~C.}\ \bibnamefont
  {Fu}}, \bibinfo {author} {\bibfnamefont {C.~W.}\ \bibnamefont {Wolgemuth}}, \
  and\ \bibinfo {author} {\bibfnamefont {T.~R.}\ \bibnamefont {Powers}},\
  }\href {\doibase 10.1103/PhysRevE.78.041913} {\bibfield  {journal} {\bibinfo
  {journal} {Phys. Rev. E}\ }\textbf {\bibinfo {volume} {78}},\ \bibinfo
  {pages} {041913} (\bibinfo {year} {2008})}\BibitemShut {NoStop}%
\bibitem [{\citenamefont {Qian}\ \emph {et~al.}(2008)\citenamefont {Qian},
  \citenamefont {Powers},\ and\ \citenamefont {Breuer}}]{Qian2008}%
  \BibitemOpen
  \bibfield  {author} {\bibinfo {author} {\bibfnamefont {B.}~\bibnamefont
  {Qian}}, \bibinfo {author} {\bibfnamefont {T.~R.}\ \bibnamefont {Powers}}, \
  and\ \bibinfo {author} {\bibfnamefont {K.~S.}\ \bibnamefont {Breuer}},\
  }\href {\doibase 10.1103/PhysRevLett.100.078101} {\bibfield  {journal}
  {\bibinfo  {journal} {Phys. Rev. Lett.}\ }\textbf {\bibinfo {volume} {100}},\
  \bibinfo {pages} {078101} (\bibinfo {year} {2008})}\BibitemShut {NoStop}%
\bibitem [{\citenamefont {Shelley}\ and\ \citenamefont
  {Zhang}(2011)}]{shelley11}%
  \BibitemOpen
  \bibfield  {author} {\bibinfo {author} {\bibfnamefont {M.~J.}\ \bibnamefont
  {Shelley}}\ and\ \bibinfo {author} {\bibfnamefont {J.}~\bibnamefont
  {Zhang}},\ }\href {\doibase 10.1146/annurev-fluid-121108-145456} {\bibfield
  {journal} {\bibinfo  {journal} {Annu. Rev. Fluid Mech.}\ }\textbf {\bibinfo
  {volume} {43}},\ \bibinfo {pages} {449} (\bibinfo {year} {2011})}\BibitemShut
  {NoStop}%
\bibitem [{\citenamefont {Dewey}\ \emph {et~al.}(2013)\citenamefont {Dewey},
  \citenamefont {Boschitsch}, \citenamefont {Moored}, \citenamefont {Stone},\
  and\ \citenamefont {Smits}}]{dewey2013}%
  \BibitemOpen
  \bibfield  {author} {\bibinfo {author} {\bibfnamefont {P.}~\bibnamefont
  {Dewey}}, \bibinfo {author} {\bibfnamefont {B.~M.}\ \bibnamefont
  {Boschitsch}}, \bibinfo {author} {\bibfnamefont {K.~W.}\ \bibnamefont
  {Moored}}, \bibinfo {author} {\bibfnamefont {H.~A.}\ \bibnamefont {Stone}}, \
  and\ \bibinfo {author} {\bibfnamefont {A.~J.}\ \bibnamefont {Smits}},\ }\href
  {\doibase 10.1017/jfm.2013.384} {\bibfield  {journal} {\bibinfo  {journal}
  {J. Fluid Mech.}\ }\textbf {\bibinfo {volume} {732}},\ \bibinfo {pages} {29}
  (\bibinfo {year} {2013})}\BibitemShut {NoStop}%
\bibitem [{\citenamefont {Wu}(2011)}]{wu11}%
  \BibitemOpen
  \bibfield  {author} {\bibinfo {author} {\bibfnamefont {T.~Y.}\ \bibnamefont
  {Wu}},\ }\href {\doibase 10.1146/annurev-fluid-122109-160648} {\bibfield
  {journal} {\bibinfo  {journal} {Annu. Rev. Fluid Mech.}\ }\textbf {\bibinfo
  {volume} {43}},\ \bibinfo {pages} {25} (\bibinfo {year} {2011})}\BibitemShut
  {NoStop}%
\bibitem [{\citenamefont {Tanaka}\ \emph {et~al.}(2011)\citenamefont {Tanaka},
  \citenamefont {Whitney},\ and\ \citenamefont {Wood}}]{tanaka2011effect}%
  \BibitemOpen
  \bibfield  {author} {\bibinfo {author} {\bibfnamefont {H.}~\bibnamefont
  {Tanaka}}, \bibinfo {author} {\bibfnamefont {J.~P.}\ \bibnamefont {Whitney}},
  \ and\ \bibinfo {author} {\bibfnamefont {R.~J.}\ \bibnamefont {Wood}},\
  }\href {\doibase 10.1093/icb/icr051} {\bibfield  {journal} {\bibinfo
  {journal} {Integr. Comp. Biol.}\ }\textbf {\bibinfo {volume} {51}},\ \bibinfo
  {pages} {142} (\bibinfo {year} {2011})}\BibitemShut {NoStop}%
\bibitem [{\citenamefont {Song}\ \emph {et~al.}(2015)\citenamefont {Song},
  \citenamefont {Luo},\ and\ \citenamefont {Hedrick}}]{song2015}%
  \BibitemOpen
  \bibfield  {author} {\bibinfo {author} {\bibfnamefont {J.}~\bibnamefont
  {Song}}, \bibinfo {author} {\bibfnamefont {H.}~\bibnamefont {Luo}}, \ and\
  \bibinfo {author} {\bibfnamefont {T.~L.}\ \bibnamefont {Hedrick}},\ }\href
  {\doibase 10.1088/1748-3190/10/1/016007} {\bibfield  {journal} {\bibinfo
  {journal} {Bioinspir. Biomim.}\ }\textbf {\bibinfo {volume} {10}},\ \bibinfo
  {pages} {016007} (\bibinfo {year} {2015})}\BibitemShut {NoStop}%
\bibitem [{\citenamefont {Shoele}\ and\ \citenamefont
  {Zhu}(2013)}]{shoele2013pof}%
  \BibitemOpen
  \bibfield  {author} {\bibinfo {author} {\bibfnamefont {K.}~\bibnamefont
  {Shoele}}\ and\ \bibinfo {author} {\bibfnamefont {Q.}~\bibnamefont {Zhu}},\
  }\href {\doibase 10.1063/1.4802193} {\bibfield  {journal} {\bibinfo
  {journal} {Phys. Fluids}\ }\textbf {\bibinfo {volume} {25}},\ \bibinfo
  {pages} {041901} (\bibinfo {year} {2013})}\BibitemShut {NoStop}%
\bibitem [{\citenamefont {Moore}(2015)}]{moore2015torsional}%
  \BibitemOpen
  \bibfield  {author} {\bibinfo {author} {\bibfnamefont {M.~N.~J.}\
  \bibnamefont {Moore}},\ }\href {\doibase 10.1063/1.4930235} {\bibfield
  {journal} {\bibinfo  {journal} {Phys. Fluids}\ }\textbf {\bibinfo {volume}
  {27}},\ \bibinfo {pages} {091701} (\bibinfo {year} {2015})}\BibitemShut
  {NoStop}%
\bibitem [{\citenamefont {Lucas}\ \emph {et~al.}(2015)\citenamefont {Lucas},
  \citenamefont {Thornycroft}, \citenamefont {Gemmell}, \citenamefont {Colin},
  \citenamefont {Costello},\ and\ \citenamefont {Lauder}}]{lucas2015effects}%
  \BibitemOpen
  \bibfield  {author} {\bibinfo {author} {\bibfnamefont {K.~N.}\ \bibnamefont
  {Lucas}}, \bibinfo {author} {\bibfnamefont {P.~J.}\ \bibnamefont
  {Thornycroft}}, \bibinfo {author} {\bibfnamefont {B.~J.}\ \bibnamefont
  {Gemmell}}, \bibinfo {author} {\bibfnamefont {S.~P.}\ \bibnamefont {Colin}},
  \bibinfo {author} {\bibfnamefont {J.~H.}\ \bibnamefont {Costello}}, \ and\
  \bibinfo {author} {\bibfnamefont {G.~V.}\ \bibnamefont {Lauder}},\ }\href
  {\doibase 10.1088/1748-3190/10/5/056019} {\bibfield  {journal} {\bibinfo
  {journal} {Bioinspir. Biomim.}\ }\textbf {\bibinfo {volume} {10}},\ \bibinfo
  {pages} {056019} (\bibinfo {year} {2015})}\BibitemShut {NoStop}%
\bibitem [{\citenamefont {Landau}\ and\ \citenamefont
  {Lifshitz}(1986)}]{Landau}%
  \BibitemOpen
  \bibfield  {author} {\bibinfo {author} {\bibfnamefont {L.~D.}\ \bibnamefont
  {Landau}}\ and\ \bibinfo {author} {\bibfnamefont {E.~M.}\ \bibnamefont
  {Lifshitz}},\ }\href@noop {} {\emph {\bibinfo {title} {Theory of
  Elasticity}}},\ \bibinfo {edition} {3rd}\ ed.\ (\bibinfo  {publisher}
  {Pergamon Press, Oxford},\ \bibinfo {year} {1986})\BibitemShut {NoStop}%
\bibitem [{\citenamefont {Goldstein}\ and\ \citenamefont
  {Langer}(1995)}]{Goldstein1995PRL}%
  \BibitemOpen
  \bibfield  {author} {\bibinfo {author} {\bibfnamefont {R.~E.}\ \bibnamefont
  {Goldstein}}\ and\ \bibinfo {author} {\bibfnamefont {S.~A.}\ \bibnamefont
  {Langer}},\ }\href {\doibase 10.1103/PhysRevLett.75.1094} {\bibfield
  {journal} {\bibinfo  {journal} {Phys. Rev. Lett.}\ }\textbf {\bibinfo
  {volume} {75}},\ \bibinfo {pages} {1094} (\bibinfo {year}
  {1995})}\BibitemShut {NoStop}%
\bibitem [{\citenamefont {Camalet}\ \emph {et~al.}(1999)\citenamefont
  {Camalet}, \citenamefont {J\"ulicher},\ and\ \citenamefont
  {Prost}}]{camalet1999PRL}%
  \BibitemOpen
  \bibfield  {author} {\bibinfo {author} {\bibfnamefont {S.}~\bibnamefont
  {Camalet}}, \bibinfo {author} {\bibfnamefont {F.}~\bibnamefont {J\"ulicher}},
  \ and\ \bibinfo {author} {\bibfnamefont {J.}~\bibnamefont {Prost}},\ }\href
  {\doibase 10.1103/PhysRevLett.82.1590} {\bibfield  {journal} {\bibinfo
  {journal} {Phys. Rev. Lett.}\ }\textbf {\bibinfo {volume} {82}},\ \bibinfo
  {pages} {1590} (\bibinfo {year} {1999})}\BibitemShut {NoStop}%
\bibitem [{\citenamefont {Camalet}\ and\ \citenamefont
  {J{\"u}licher}(2000)}]{camalet2000}%
  \BibitemOpen
  \bibfield  {author} {\bibinfo {author} {\bibfnamefont {S.}~\bibnamefont
  {Camalet}}\ and\ \bibinfo {author} {\bibfnamefont {F.}~\bibnamefont
  {J{\"u}licher}},\ }\href {\doibase 10.1088/1367-2630/2/1/324} {\bibfield
  {journal} {\bibinfo  {journal} {New J. Phys.}\ }\textbf {\bibinfo {volume}
  {2}},\ \bibinfo {pages} {24} (\bibinfo {year} {2000})}\BibitemShut {NoStop}%
\bibitem [{\citenamefont {Yu}\ \emph {et~al.}(2006)\citenamefont {Yu},
  \citenamefont {Lauga},\ and\ \citenamefont {Hosoi}}]{Yu2006}%
  \BibitemOpen
  \bibfield  {author} {\bibinfo {author} {\bibfnamefont {T.~S.}\ \bibnamefont
  {Yu}}, \bibinfo {author} {\bibfnamefont {E.}~\bibnamefont {Lauga}}, \ and\
  \bibinfo {author} {\bibfnamefont {A.~E.}\ \bibnamefont {Hosoi}},\ }\href
  {\doibase 10.1063/1.2349585} {\bibfield  {journal} {\bibinfo  {journal}
  {Phys. Fluids}\ }\textbf {\bibinfo {volume} {18}},\ \bibinfo {pages} {091701}
  (\bibinfo {year} {2006})}\BibitemShut {NoStop}%
\bibitem [{\citenamefont {Evans}\ and\ \citenamefont
  {Lauga}(2010)}]{Evans2010PRE}%
  \BibitemOpen
  \bibfield  {author} {\bibinfo {author} {\bibfnamefont {A.~A.}\ \bibnamefont
  {Evans}}\ and\ \bibinfo {author} {\bibfnamefont {E.}~\bibnamefont {Lauga}},\
  }\href {\doibase 10.1103/PhysRevE.82.041915} {\bibfield  {journal} {\bibinfo
  {journal} {Phys. Rev. E}\ }\textbf {\bibinfo {volume} {82}},\ \bibinfo
  {pages} {041915} (\bibinfo {year} {2010})}\BibitemShut {NoStop}%
\bibitem [{\citenamefont {Spagnolie}\ and\ \citenamefont
  {Lauga}(2010)}]{spagnolie2010optimal}%
  \BibitemOpen
  \bibfield  {author} {\bibinfo {author} {\bibfnamefont {S.~E.}\ \bibnamefont
  {Spagnolie}}\ and\ \bibinfo {author} {\bibfnamefont {E.}~\bibnamefont
  {Lauga}},\ }\href {\doibase 10.1063/1.3318497} {\bibfield  {journal}
  {\bibinfo  {journal} {Phys. Fluids}\ }\textbf {\bibinfo {volume} {22}},\
  \bibinfo {pages} {031901} (\bibinfo {year} {2010})}\BibitemShut {NoStop}%
\bibitem [{\citenamefont {Riedel-Kruse}\ \emph {et~al.}(2007)\citenamefont
  {Riedel-Kruse}, \citenamefont {Hilfinger}, \citenamefont {Howard},\ and\
  \citenamefont {J\"ulicher}}]{rk_HFSP_J}%
  \BibitemOpen
  \bibfield  {author} {\bibinfo {author} {\bibfnamefont {I.~H.}\ \bibnamefont
  {Riedel-Kruse}}, \bibinfo {author} {\bibfnamefont {A.}~\bibnamefont
  {Hilfinger}}, \bibinfo {author} {\bibfnamefont {J.}~\bibnamefont {Howard}}, \
  and\ \bibinfo {author} {\bibfnamefont {F.}~\bibnamefont {J\"ulicher}},\
  }\href {\doibase 10.2976/1.2773861} {\bibfield  {journal} {\bibinfo
  {journal} {HFSP J.}\ }\textbf {\bibinfo {volume} {1}},\ \bibinfo {pages}
  {192} (\bibinfo {year} {2007})}\BibitemShut {NoStop}%
\bibitem [{\citenamefont {Gray}\ and\ \citenamefont
  {Hancock}(1955)}]{Gray1955}%
  \BibitemOpen
  \bibfield  {author} {\bibinfo {author} {\bibfnamefont {J.}~\bibnamefont
  {Gray}}\ and\ \bibinfo {author} {\bibfnamefont {G.~J.}\ \bibnamefont
  {Hancock}},\ }\href@noop {} {\bibfield  {journal} {\bibinfo  {journal} {J.
  Exp. Biol.}\ }\textbf {\bibinfo {volume} {32}},\ \bibinfo {pages} {802}
  (\bibinfo {year} {1955})}\BibitemShut {NoStop}%
\bibitem [{\citenamefont {Lighthill}(1975)}]{lighthill1975mathematica}%
  \BibitemOpen
  \bibfield  {author} {\bibinfo {author} {\bibfnamefont {J.}~\bibnamefont
  {Lighthill}},\ }\href@noop {} {\emph {\bibinfo {title} {Mathematical
  Biofluiddynamics}}}\ (\bibinfo  {publisher} {SIAM, Philadelphia},\ \bibinfo
  {year} {1975})\BibitemShut {NoStop}%
\bibitem [{\citenamefont {Childress}(1981)}]{childress1981mechanics}%
  \BibitemOpen
  \bibfield  {author} {\bibinfo {author} {\bibfnamefont {S.}~\bibnamefont
  {Childress}},\ }\href@noop {} {\emph {\bibinfo {title} {Mechanics of swimming
  and flying}}}\ (\bibinfo  {publisher} {Cambridge University Press},\ \bibinfo
  {year} {1981})\BibitemShut {NoStop}%
\bibitem [{\citenamefont {Audoly}\ and\ \citenamefont
  {Pomeau}(2010)}]{audoly2010elasticity}%
  \BibitemOpen
  \bibfield  {author} {\bibinfo {author} {\bibfnamefont {B.}~\bibnamefont
  {Audoly}}\ and\ \bibinfo {author} {\bibfnamefont {Y.}~\bibnamefont
  {Pomeau}},\ }\href@noop {} {\emph {\bibinfo {title} {Elasticity and geometry:
  from hair curls to the non-linear response of shells}}}\ (\bibinfo
  {publisher} {Oxford University Press},\ \bibinfo {year} {2010})\BibitemShut
  {NoStop}%
\bibitem [{\citenamefont {Pak}\ and\ \citenamefont
  {Lauga}(2014)}]{pak2014theoretical}%
  \BibitemOpen
  \bibfield  {author} {\bibinfo {author} {\bibfnamefont {O.~S.}\ \bibnamefont
  {Pak}}\ and\ \bibinfo {author} {\bibfnamefont {E.}~\bibnamefont {Lauga}},\
  }in\ \href@noop {} {\emph {\bibinfo {booktitle} {Fluid-Structure Interactions
  in Low-Reynolds-Number Flows}}},\ \bibinfo {editor} {edited by\ \bibinfo
  {editor} {\bibfnamefont {C.}~\bibnamefont {Duprat}}\ and\ \bibinfo {editor}
  {\bibfnamefont {H.~A.}\ \bibnamefont {Stone}}}\ (\bibinfo  {publisher} {Royal
  Society of Chemistry},\ \bibinfo {year} {2014})\ p.\ \bibinfo {pages}
  {100}\BibitemShut {NoStop}%
\bibitem [{\citenamefont {Gao}\ \emph {et~al.}(2012)\citenamefont {Gao},
  \citenamefont {Kagan}, \citenamefont {Pak}, \citenamefont {Clawson},
  \citenamefont {Campuzano}, \citenamefont {Chuluun-Erdene}, \citenamefont
  {Shipton}, \citenamefont {Fullerton}, \citenamefont {Zhang}, \citenamefont
  {Lauga},\ and\ \citenamefont {Wang}}]{Gao2012}%
  \BibitemOpen
  \bibfield  {author} {\bibinfo {author} {\bibfnamefont {W.}~\bibnamefont
  {Gao}}, \bibinfo {author} {\bibfnamefont {D.}~\bibnamefont {Kagan}}, \bibinfo
  {author} {\bibfnamefont {O.~S.}\ \bibnamefont {Pak}}, \bibinfo {author}
  {\bibfnamefont {C.}~\bibnamefont {Clawson}}, \bibinfo {author} {\bibfnamefont
  {S.}~\bibnamefont {Campuzano}}, \bibinfo {author} {\bibfnamefont
  {E.}~\bibnamefont {Chuluun-Erdene}}, \bibinfo {author} {\bibfnamefont
  {E.}~\bibnamefont {Shipton}}, \bibinfo {author} {\bibfnamefont {E.~E.}\
  \bibnamefont {Fullerton}}, \bibinfo {author} {\bibfnamefont {L.}~\bibnamefont
  {Zhang}}, \bibinfo {author} {\bibfnamefont {E.}~\bibnamefont {Lauga}}, \ and\
  \bibinfo {author} {\bibfnamefont {J.}~\bibnamefont {Wang}},\ }\href {\doibase
  10.1002/smll.201101909} {\bibfield  {journal} {\bibinfo  {journal} {Small}\
  }\textbf {\bibinfo {volume} {8}},\ \bibinfo {pages} {460} (\bibinfo {year}
  {2012})}\BibitemShut {NoStop}%
\bibitem [{\citenamefont {Schamel}\ \emph {et~al.}(2014)\citenamefont
  {Schamel}, \citenamefont {Mark}, \citenamefont {Gibbs}, \citenamefont
  {Miksch}, \citenamefont {Morozov}, \citenamefont {Leshansky},\ and\
  \citenamefont {Fischer}}]{Schamel2014}%
  \BibitemOpen
  \bibfield  {author} {\bibinfo {author} {\bibfnamefont {D.}~\bibnamefont
  {Schamel}}, \bibinfo {author} {\bibfnamefont {A.~G.}\ \bibnamefont {Mark}},
  \bibinfo {author} {\bibfnamefont {J.~G.}\ \bibnamefont {Gibbs}}, \bibinfo
  {author} {\bibfnamefont {C.}~\bibnamefont {Miksch}}, \bibinfo {author}
  {\bibfnamefont {K.~I.}\ \bibnamefont {Morozov}}, \bibinfo {author}
  {\bibfnamefont {A.~M.}\ \bibnamefont {Leshansky}}, \ and\ \bibinfo {author}
  {\bibfnamefont {P.}~\bibnamefont {Fischer}},\ }\href {\doibase
  10.1021/nn502360t} {\bibfield  {journal} {\bibinfo  {journal} {ACS Nano}\
  }\textbf {\bibinfo {volume} {8}},\ \bibinfo {pages} {8794} (\bibinfo {year}
  {2014})}\BibitemShut {NoStop}%
\bibitem [{\citenamefont {Cheang}\ and\ \citenamefont
  {Kim}(2015)}]{Cheang2015}%
  \BibitemOpen
  \bibfield  {author} {\bibinfo {author} {\bibfnamefont {U.~K.}\ \bibnamefont
  {Cheang}}\ and\ \bibinfo {author} {\bibfnamefont {M.~J.}\ \bibnamefont
  {Kim}},\ }\href {\doibase 10.1007/s11051-014-2737-z} {\bibfield  {journal}
  {\bibinfo  {journal} {J. Nanopart. Res.}\ }\textbf {\bibinfo {volume} {17}},\
  \bibinfo {pages} {1} (\bibinfo {year} {2015})}\BibitemShut {NoStop}%
\bibitem [{\citenamefont {Shaijumon}\ \emph {et~al.}(2008)\citenamefont
  {Shaijumon}, \citenamefont {Ou}, \citenamefont {Ci},\ and\ \citenamefont
  {Ajayan}}]{Shaijumon08}%
  \BibitemOpen
  \bibfield  {author} {\bibinfo {author} {\bibfnamefont {M.~M.}\ \bibnamefont
  {Shaijumon}}, \bibinfo {author} {\bibfnamefont {F.~S.}\ \bibnamefont {Ou}},
  \bibinfo {author} {\bibfnamefont {L.}~\bibnamefont {Ci}}, \ and\ \bibinfo
  {author} {\bibfnamefont {P.~M.}\ \bibnamefont {Ajayan}},\ }\href {\doibase
  10.1039/B800866C} {\bibfield  {journal} {\bibinfo  {journal} {Chem. Commun.}\
  ,\ \bibinfo {pages} {2373}} (\bibinfo {year} {2008})}\BibitemShut {NoStop}%
\bibitem [{\citenamefont {Kitzhofer}\ \emph {et~al.}(2010)\citenamefont
  {Kitzhofer}, \citenamefont {Koch}, \citenamefont {Pulverer}, \citenamefont
  {Simon},\ and\ \citenamefont {Weinm{\"u}ller}}]{kitzhofer2010new}%
  \BibitemOpen
  \bibfield  {author} {\bibinfo {author} {\bibfnamefont {G.}~\bibnamefont
  {Kitzhofer}}, \bibinfo {author} {\bibfnamefont {O.}~\bibnamefont {Koch}},
  \bibinfo {author} {\bibfnamefont {G.}~\bibnamefont {Pulverer}}, \bibinfo
  {author} {\bibfnamefont {C.}~\bibnamefont {Simon}}, \ and\ \bibinfo {author}
  {\bibfnamefont {E.}~\bibnamefont {Weinm{\"u}ller}},\ }\href@noop {}
  {\bibfield  {journal} {\bibinfo  {journal} {J. Numer. Anal. Indust. Appl.
  Math}\ }\textbf {\bibinfo {volume} {5}},\ \bibinfo {pages} {113} (\bibinfo
  {year} {2010})}\BibitemShut {NoStop}%
\bibitem [{\citenamefont {De~Hoog}\ and\ \citenamefont
  {Weiss}(1976)}]{de1976difference}%
  \BibitemOpen
  \bibfield  {author} {\bibinfo {author} {\bibfnamefont {F.~R.}\ \bibnamefont
  {De~Hoog}}\ and\ \bibinfo {author} {\bibfnamefont {R.}~\bibnamefont
  {Weiss}},\ }\href {\doibase 10.1137/0713063} {\bibfield  {journal} {\bibinfo
  {journal} {SIAM J. Numer. Anal.}\ }\textbf {\bibinfo {volume} {13}},\
  \bibinfo {pages} {775} (\bibinfo {year} {1976})}\BibitemShut {NoStop}%
\bibitem [{\citenamefont {de~Hoog}\ and\ \citenamefont
  {Weiss}(1980)}]{de1980boundary}%
  \BibitemOpen
  \bibfield  {author} {\bibinfo {author} {\bibfnamefont {F.~R.}\ \bibnamefont
  {de~Hoog}}\ and\ \bibinfo {author} {\bibfnamefont {R.}~\bibnamefont
  {Weiss}},\ }\href {\doibase 10.1137/0511003} {\bibfield  {journal} {\bibinfo
  {journal} {SIAM J. Math. Anal.}\ }\textbf {\bibinfo {volume} {11}},\ \bibinfo
  {pages} {41} (\bibinfo {year} {1980})}\BibitemShut {NoStop}%
\end{thebibliography}%
